\documentclass[12pt]{article}
\usepackage[english]{babel}
\usepackage[latin1]{inputenc}
\usepackage{amsfonts,amssymb,amsmath, epsfig}
\usepackage{color,graphicx,graphics,psfrag}
\usepackage{amsmath,amstext,amssymb,amsfonts, amscd}
\usepackage[lofdepth,lotdepth]{subfig}
\usepackage{multirow}
\usepackage{hyperref}          

\def\Box{\leavevmode\vbox{\hrule
     \hbox{\vrule\kern4pt\vbox{\kern4pt}%
           \vrule}\hrule}}
\def\blackbox{\leavevmode\vrule height 5pt width 4pt depth 0pt\relax}
\def\endproof{\null\hfill {$\blackbox$}\bigskip}

\def\paragraph#1{{\bf #1\ }}

\newtheorem{lemma}{Lemma}[section]

\newtheorem{definition}[lemma]{Definition}

\newtheorem{proposition}[lemma]{Proposition}

\newtheorem{remark}{Remark}[section]

\title{Topological travelling waves of a macroscopic swarmalator model in confined geometries} 

\author{P. Degond$^{(1)}$, A. Diez$^{(2)}$} 
\date{} 
\begin{document}

\maketitle

\vspace{0.5 cm}

\begin{center}
$^{(1)}$ Institut de Math\'ematiques de Toulouse ; UMR5219 \\
Universit\'e de Toulouse ; CNRS \\
UPS, F-31062 Toulouse Cedex 9, France\\
email: pierre.degond@math.univ-toulouse.fr

\bigskip

$^{(2)}$ Institute for the Advanced Study of Human Biology (ASHBi) \\
Kyoto University Institute for Advanced Study, \\ 
Yoshida-Konoe-cho, Sakyo-ku, Kyoto 606-8501, Japan\\
email: diez.antoinenicolas.4e@kyoto-u.ac.jp

\vspace{0.8cm}
{\em Dedicated to Shi Jin, inspiring mathematician and friend}

\end{center}

\vspace{0.5 cm}
\begin{abstract}
We investigate a new class of topological travelling-wave solutions for a macroscopipc swarmalator model involving force non-reciprocity. Swarmalators are systems of self-propelled particles endowed with a phase variable. The particles are subject to coupled swarming and synchronization. In previous work, the swarmalator under study was introduced, the macroscopic model was derived and doubly periodic travelling-wave solutions were exhibited. Here, we focus on the macroscopic model and investigate new classes of two-dimensional travelling-wave solutions. These solutions are confined in a strip or in an annulus. In the case of the strip, they are periodic along the strip direction. They have non-trivial topology as their phase increases by a multiple of $2 \pi$ from one period (in the case of the strip) or one revolution (in the case of the annulus) to the next. Existence and qualitative behavior of these solutions are investigated.
\end{abstract}



\medskip
\noindent
{\bf Acknowledgements:} PD holds a visiting professor association with the Department ofMathematics, Imperial College London where part of this research was conducted. Part of this work was conducted when AD was affiliated to the Department of Mathematics, Imperial College London and supported by an Imperial College Roth scholarship cofunded with the Engineering and Physical Sciences Research Council.

\medskip
\noindent
{\bf Data statement:} No new data were collected in the course of this research.

\setcounter{equation}{0}
\section{Introduction}
\label{intro}

In this paper, we investigate new classes of topological travelling-wave solutions of a macroscopic swarmalator model first derived in \cite{degond2022topological}. Swarmalators are a special class of particle systems which combine the features of swarming systems and oscillators, hence the terminology first coined in \cite{o2017oscillators}. Swarming systems have been the subject of an intense literature. Among them, two classes of model have received particular attention: the Vicsek model on the one hand, which imposes a constant velocity to the agents (see e.g. the original article \cite{vicsek1995novel} and its many follow-ups \cite{aoki1982simulation, calovi2014swarming, cavagna2015flocking, chate2008collective, costanzo2018spontaneous, couzin2002collective, degond2015phase, degond2008continuum, frouvelle2012continuum} as well as the review~\cite{vicsek2012collective}) and the Cucker-Smale model which does not impose such a constraint (see the original work~\cite{cucker2007emergent} and its follow-ups  \cite{aceves2019hydrodynamic, barbaro2016phase, dorsogna2006self, ha2009simple, ha2008from, motsch2011new}). On the other hand, synchronization of oscillator systems has also stimulated a vast literature (see \cite{kuramoto2003chemical} and the review \cite{acebron2005kuramoto}). Swarmalators have recently emerged as an active research subject (see the original work \cite{o2017oscillators}, its recent elaborations \cite{degond2022topological, ha2019emergent, hong2018active, hong2021coupling, jimenez2020oscillatory, lee2021collective, lizarraga2020synchronization, o2022collective, o2018ring} and the review~\cite{o2019review}). Swarming systems have a lot of applications in the description of living systems (see e.g. \cite{boissard2013trail, bricard2015emergent, couzin2003self, creppy2016symmetry, czirok1996formation, gautrais2012deciphering, hemelrijk2010emergence, lukeman2010inferring}. Swarmalators have been specifically applied to living systems in \cite{japon2021intercellular, peshkov2022synchronized}

In this paper, we aim to investigate systems which have solutions with non-trivial topology. Topology in physical systems has recently emerged as a subject of major interest in relation to the discovery of so-called topological insulators \cite{hasan2010colloquium, qi2011topological}. Solutions with non-trivial topology show increased robustness against perturbations, a feature called ``topological protection''. Applications of topology to collective dynamics have recently caught the attention of many research groups \cite{degond2021bulk, shankar2017topological, sone2019anomalous, souslov2017topological, zhang2020oscillatory} (see also the review \cite{shankar2022topological}). Topology in swarmalator models manifests itself in the so called  ``phase-wave states'' of~\cite{o2017oscillators}.

In the present paper, we consider a macroscopic swarmalator model originally derived in  \cite{degond2022topological}. At the microscopic level, this model relies on the Vicsek model for the self-propulsion velocity dynamics i.e. velocities tend to align with the average velocity of the surrounding group of particles, up to some small noise. But the particle positions are subject both to motion along the self-propulsion velocity and to a phase-dependent attraction-repulsion force. In turn, the phases tend to align with those of the surrounding particles up to some noise. In this way, there is a bi-directional coupling between direction of self-propulsion and phase. An original feature of this model is that attraction-repulsion forces do not obey the reciprocity principle, but rather, result in pursuit behavior within pairs of particles of different phases.    

In \cite{degond2022topological}, the corresponding macroscopic model was derived from the  particle system and exploited to exhibit a class of doubly-periodic topological travelling-waves. The details of the derivation procedure, which classically involves an intermediate model between the particle and hydrodynamic ones, namely the kinetic model, can be found in \cite{degond2022topological}. We also refer to \cite{degond2022topological} for a survey of the methodologies used. In the present paper we show that this hydrodynamic model supports other classes of topological travelling-wave solutions associated with strip and annular geometry. The strip geometry leads to a quasi-explicit analytical treatment. By contrast, the annular geometry (which is classically considered in the literature see e.g. the ``phase-wave states'' of~\cite{o2017oscillators}) does not lend itself to such a straightforward treatment and requires the use of methods of more analytical and geometrical nature. A similar but one-dimensional ring geometry has been recently considered in \cite{Hong:2023aa} for the original swarmalator model of \cite{o2017oscillators}. Note that the strip case is not the limiting case of an annulus whose radii tend to infinity because in this limit the solution becomes trivial due to a normalization condition. The main features of all these solutions is that they present a non-trivial geometry as the index of the phase vector in a period is non zero. 

The organization of this paper is as follows. In Section \ref{sec_part_kin}, we introduce the model. Topological travelling-wave solutions in strip geometry are presented in Section \ref{sec:explicit_strip} and in annular geometry, in Section \ref{sec:explicit_annular}. Proofs of existence of such solutions are given in Section \ref{sec_strip_proofs} in the case of strip geometry and in Section \ref{sec_rotat_sym_proof} in the case of annular geometry. A numerical illustration for one of the classes of solutions is shown in Section \ref{sec:numerics}. Finally a conclusion and a discussion are developed in Section \ref{sec:conclusion}.

\setcounter{equation}{0}
\section{Presentation of the model}
\label{sec_part_kin}

The model under consideration has been derived in \cite{degond2022topological} and is written as follows. We consider the position variable $x \in {\mathbb R}^n$ and the time variable $t \geq 0$ and we look for a function $(\rho, u, \alpha)$: ${\mathbb R}^n \times [0,\infty) \to [0,\infty) \times {\mathbb S}^{n-1} \times {\mathbb R}/(2 \pi {\mathbb Z})$ (where ${\mathbb S}^{n-1}$ denotes the $n-1$-dimensional unit sphere), whose components respectively represent the mean density, mean self-propulsion direction and mean phase of the particles in a small fluid element at position $x$ and time $t$. They solve the following system 
\begin{eqnarray}
&&\hspace{-1cm}
\partial_t \rho + \nabla_x \cdot \big[ \rho (c_1 u + b  \rho \nabla_x \alpha) \big] = 0, \label{eq:fl_rho} \\
&&\hspace{-1cm}
\partial_t u + \big[ (c_2 u + b  \rho \nabla_x \alpha) \cdot \nabla_x \big] u + P_{u^\bot} \nabla_x ( \Theta  \log \rho + \kappa V ) = 0, \label{eq:fl_u} \\
&&\hspace{-1cm}
\rho \, \big( \partial_t \alpha + \big[ (c_1 u + b' \rho \nabla_x \alpha) \cdot \nabla_x \big] \alpha \big) - \Theta' \, \nabla_x \cdot \big( \rho \nabla_x \rho \big) = 0, \label{eq:fl_al}
\end{eqnarray}
where $c_1$, $b$, $c_2$, $\Theta$, $\kappa$, $b'$ and $\Theta'$ are real constants having the following properties: 
\begin{equation}
0 < c_2 < c_1 < 1, \qquad \kappa>0, \qquad \Theta >0. 
\label{eq:prop_c1_c2_etc}
\end{equation} 
The function ${\mathbb R}^n \ni x \mapsto V(x) \in {\mathbb R}$ is a given external confinement potential and $P_{u^\bot} = \textrm{Id} - u \otimes u$ is the projection matrix onto $\{u\}^\bot$. The density $\rho$ is normalized, i.e.
\begin{equation}
\int_{{\mathbb R}^n} \rho(x,t) \, dx = 1, 
\label{eq:rho_normaliz}
\end{equation}
because, for any $\mu >0$, $\mu \rho$ is a solution of the same system with $b$, $b'$ and $\Theta'$ changed into $\mu b$, $\mu b'$ and $\mu \Theta'$. 

This model has been referred to in \cite{degond2022topological} as ``Swarmalator Hydrodynamics'' (SH). Eq.~\eqref{eq:fl_rho} is the fluid continuity equation, where the fluid velocity $c_1 u + b  \rho \nabla_x \alpha$ has two components. The first one stems from the self-propulsion forces of the particles; it is in the direction of the mean self-propulsion direction $u$ and has magnitude $c_1$. The second one comes from the attraction-repulsion forces originating from phase mean-phase gradients~$\nabla \alpha$ with coupling intensity $b$. Eq. \eqref{eq:fl_u} describes how the mean self-propulsion direction evolves. It is passively transported by the velocity field $c_2 u + b  \rho \nabla_x \alpha$ which has the same two components as the fluid velocity except that the mean self-propulsion direction is weighted by a different coefficient $c_2$. The fact that $c_2 < c_1$ is key to the existence of travelling-wave solutions as described below. The transport of $u$ is balanced by $P_{u^\bot} \nabla_x (\Theta \log \rho + \kappa V)$. The prefactor $P_{u^\bot}$ makes sure that $u$ remains of unit norm (i.e. $u \in {\mathbb S}^{n-1}$) in the course of time. The first term is the pressure force where $\Theta$ plays the role of the fluid temperature. The second term describes the influence of the confinement potential. Finally, Eq. \eqref{eq:fl_al} describes the evolution of the mean particle phase. Like for $u$, the phase $\alpha$ is passively transported by the velocity field $c_1 u + b'  \rho \nabla_x \alpha$ which has the same two components as the fluid velocity except that now the influence of the phase gradient has a different weight $b'$. The transport of the phase is balanced by a term which describes the influence of first- and second-order derivatives of the density. 

More detailed comments can be found in \cite{degond2022topological}. All model coefficients are related to corresponding coefficients of the particle model. In particular, it is shown that in some regime (infinitesimally small noise in the evolution of the particle phases), we can make the following simplifications: 
\begin{equation}
b=b'= - \gamma, \quad \Theta'=0. 
\label{eq:simplific_small_noise}
\end{equation}
Then, the model reduces to: 
\begin{eqnarray}
&&\hspace{-1cm}
\partial_t \rho + \nabla_x \cdot \big[ \rho (c_1 u + b  \rho \nabla_x \alpha) \big] = 0, \label{eq:fl_rho_sm} \\
&&\hspace{-1cm}
\partial_t u + \big[ (c_2 u  + b  \rho \nabla_x \alpha) \cdot \nabla_x \big] u + P_{u^\bot}  \nabla_x ( \Theta \log \rho  + \kappa V ) = 0, \label{eq:fl_u_sm} \\
&&\hspace{-1cm}
\partial_t \alpha + \big[ (c_1 u  + b \rho \nabla_x \alpha) \cdot \nabla_x \big] \alpha = 0, \label{eq:fl_al_sm}
\end{eqnarray}
together with the normalization condition \eqref{eq:rho_normaliz}, and was referred to in \cite{degond2022topological} as the ``Noiseless Swarmalator Hydrodynamics'' (NSH). This model will be the main focus of the present paper. It can be rephrased as a constrained system of first order partial differential equations for the unknowns $(\rho, u, \nabla_x \alpha)$. This system has been shown to be hyperbolic about uniform states  $(\rho_0, u_0, \nabla_x \alpha_0)$ such that $u_0$ and $\nabla_x \alpha_0$ are aligned but the hyperbolicity may be lost when $u_0$ and $\nabla_x \alpha_0$ are not aligned \cite[Lemma 3.6]{degond2022topological}.

\setcounter{equation}{0}
\section{Travelling-wave solutions in strip geometry}
\label{sec:explicit_strip}

\subsection{Setting}
\label{subsec:setting}
 
In this section, we restrict ourselves to dimension $n=2$. We let $(x_1,x_2)$ be the cartesian coordinates of a point $x \in {\mathbb R}^2$ and denote by $(e_1,e_2)$ the cartesian coordinate basis and by $(u_1,u_2)$ the two coordinates of the self-propulsion velocity $u$. We recall that $u$ is a normalized vector, i.e. 
\begin{equation}
u_1^2 + u_2^2 = 1.  
\label{eq:str_const}
\end{equation}
In the previous paper \cite{degond2022topological}, doubly-periodic travelling-wave solutions were exhibited. Here, we will derive travelling-wave solutions in strip geometry. We assume a spatial domain $\Omega = (-1/2,1/2) \times (0,1)$ with periodic boundary conditions with respect to $x_2$. We let $V = V(x_1)$ be defined for $x_1 \in (-1/2,1/2)$. We assume that $V$ is smooth, even, strictly convex with $V(0)=0$, $\partial_{x_1}^2 V(0) >0$ and $V(x) \to + \infty$ when  $|x| \to 1/2$. 

We let $Z \in 2 \pi {\mathbb Z} \setminus \{0\}$ and $\lambda \in {\mathbb R}$ and define 
\begin{equation}
\ell =: - \frac{\lambda}{bZ}. 
\label{eq:def_lam}
\end{equation}
We introduce the following notations: 
\begin{equation} 
q = \frac{\Theta}{\Theta + c_1-c_2} \in (0,1), \quad \kappa' = \frac{\kappa}{\Theta + c_1-c_2}, \quad
{\mathcal I} = \int_{-1/2}^{1/2} e^{- \kappa' V(x_1)} \, d x_1,  \label{eq:def_calI} 
\end{equation}
where we recall that we assume condition \eqref{eq:prop_c1_c2_etc}. The following function
\begin{equation}
F: \quad [0,\infty) \times {\mathbb R} \to [0,\infty), \qquad (\rho, \ell) \mapsto F(\rho,\ell) = \rho^q \, |\rho + \ell|^{1-q}, 
\label{eq:definition_of_F}
\end{equation}
will be central to this study. 

If $\ell \geq 0$, $F(\cdot, \ell)$: $[0,\infty) \to [0,\infty)$, $\rho \mapsto F(\rho, \ell)$  is continuous and strictly increasing and we denote by $G(\cdot, \ell)$ its inverse. Since $F(\rho, \ell) \sim \rho$ when $\rho \to +\infty$, $F(\cdot, \ell)$ is onto $[0,\infty)$. Hence, we have $G$: $[0,\infty)^2 \to [0,\infty)$, $(y,\ell) \mapsto G(y, \ell)$. Moreover, $F$ and consequently, $G$ belong to $C^0\big([0,\infty)^2 \big) \cap C^\infty\big((0,\infty)^2 \big)$.

\begin{figure}[htbp]
\centering

$\mbox{}$ \vspace{-0.2cm} \hspace{0.cm}
\subfloat[]{\includegraphics[trim={3.5cm 13cm 5.cm 4.5cm},clip,height= 6.1cm]{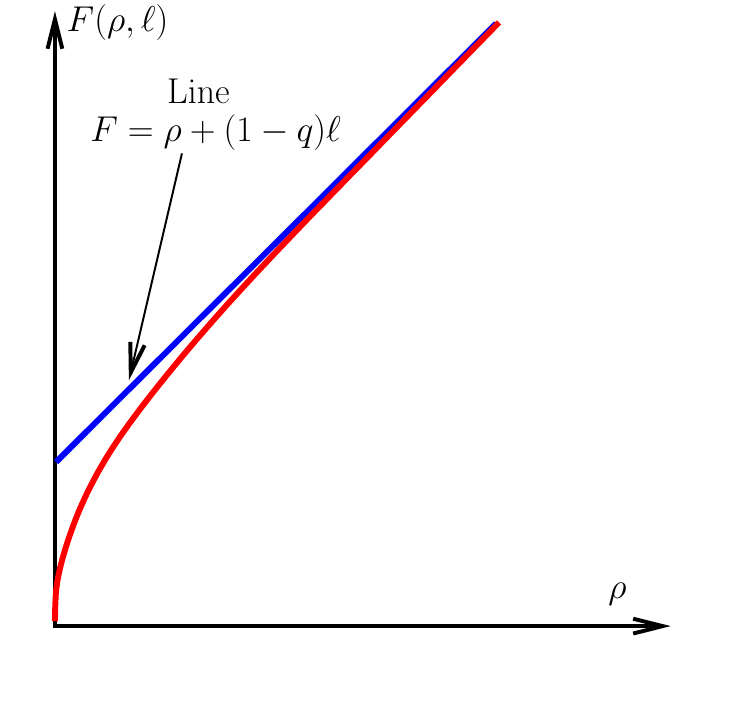}\label{subfig:function_F}} \hspace{0.5cm}
\subfloat[]{\includegraphics[trim={3.5cm 13cm 5.cm 4.5cm},clip,height= 6.1cm]{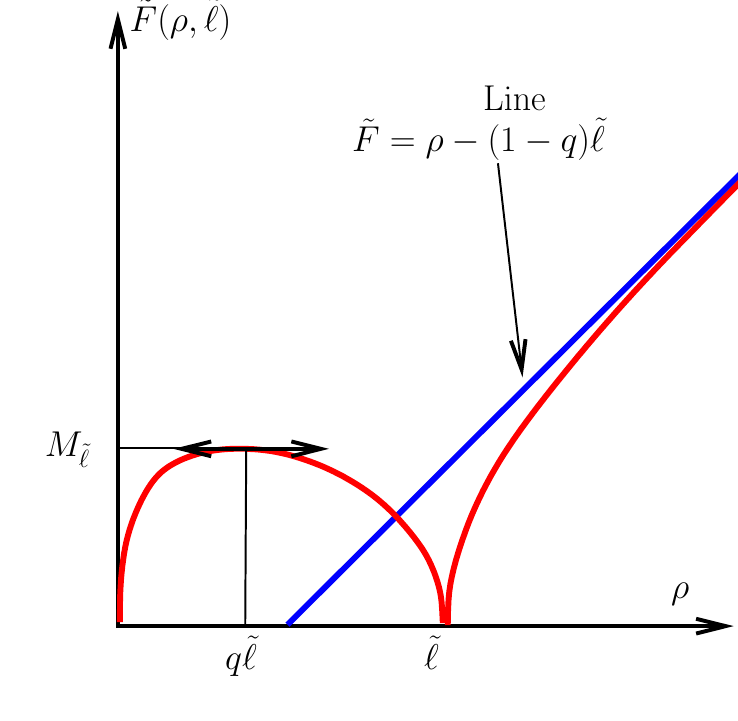}\label{subfig:function_tilF}} 

$\mbox{}$ \vspace{-0.2cm} \hspace{0.1cm}
\subfloat[]{\includegraphics[trim={3.5cm 13.cm 5.cm 4.5cm},clip,height= 6.1cm]{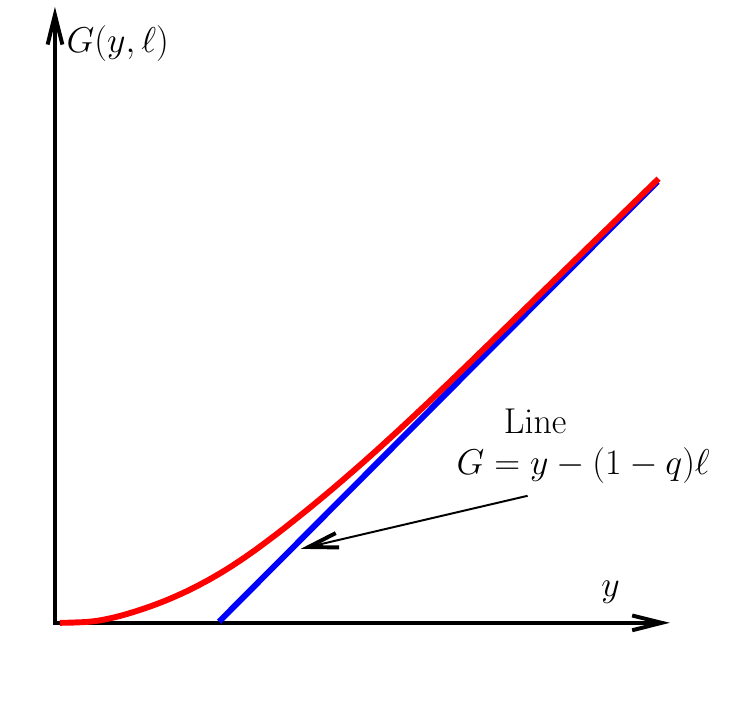}\label{subfig:function_G}} \hspace{0.5cm}
\subfloat[]{\includegraphics[trim={3.5cm 13.cm 5.cm 4.5cm},clip,height= 6.1cm]{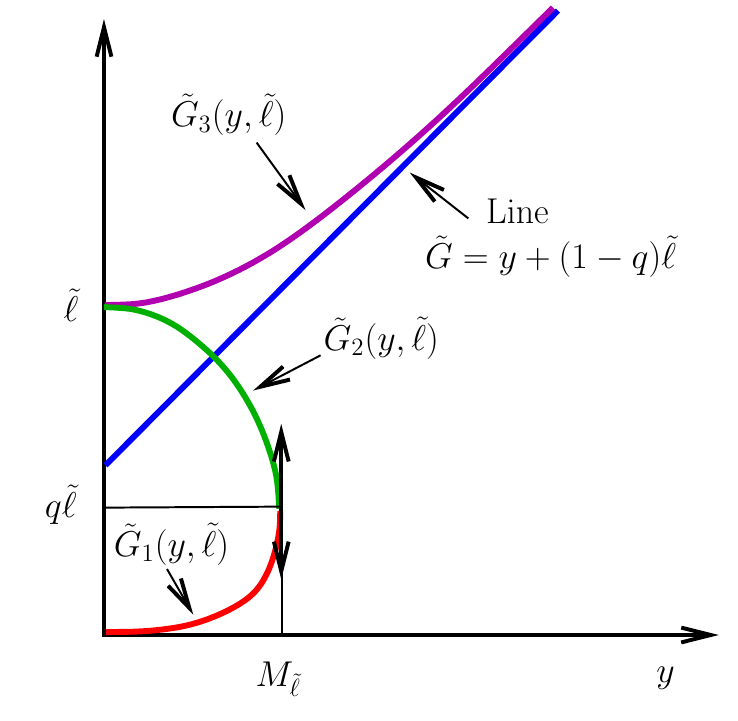}\label{subfig:function_tilG}}

\caption{(a) Function $\rho \mapsto F(\rho,\ell)$ for $\ell >0$. (b) Function $\rho \mapsto \tilde F(\rho,\tilde \ell)$ for $\tilde \ell = - \ell >0$. (c) Function $y \mapsto G(y,\ell)$ for $\ell >0$. (d) Functions $y \mapsto \tilde G_k(y,\tilde \ell)$ for $k=1, \, 2, \, 3$ and $\tilde \ell = - \ell >0$. $\tilde G_1$, $\tilde G_2$ and $\tilde G_3$ are depicted in red, green and magenta respectively.}
\label{fig:plotFG}
\end{figure}

If $\ell < 0$, we will denote $\tilde \ell = - \ell >0$ and $\tilde F(\rho, \tilde \ell) = F(\rho, \ell)$. We have: 
$$ \tilde F(\rho, \tilde \ell) = \left\{ \begin{array}{lll} \rho^q (\tilde \ell - \rho)^{1-q} & \textrm{if} & \rho \leq \tilde \ell, \\
\rho^q (\rho - \tilde \ell)^{1-q} & \textrm{if} & \rho \geq \tilde \ell.  \end{array} \right. 
$$
$\tilde F(\cdot, \ell)$ is increasing in the interval $[0, q \tilde \ell]$, decreasing in $[q \tilde \ell, \tilde \ell]$ and increasing again in $[\tilde \ell, \infty)$. It reaches a local maximum at $q \tilde \ell$ with 
\begin{equation}
\tilde F(q \tilde \ell, \tilde \ell) = M_{\tilde \ell} =: m_q \tilde \ell, \quad \textrm{ with } \quad m_q = q^q (1-q)^{1-q}.
\label{eq:def_Mtill}
\end{equation}
Thus, we can define three inverses: $G_1(\cdot, \tilde \ell)$: $[0,M_{\tilde \ell}] \to [0,q \tilde \ell]$, $G_2(\cdot, \tilde \ell)$: $[0,M_{\tilde \ell}] \to [q \tilde \ell,\tilde \ell]$ and $G_3(\cdot, \tilde \ell)$: $[0,\infty) \to [\tilde \ell,\infty)$. 
$G_1(\cdot, \tilde \ell)$ and $G_3(\cdot, \tilde \ell)$ are increasing while $G_2(\cdot, \tilde \ell)$ is decreasing. They are all continuous on their domains of definition and $C^\infty$ in their interior. We refer to Fig. \ref{fig:plotFG} for a plot of the functions $F(\cdot, \ell)$ and $\tilde F(\cdot, \tilde \ell)$ as well as their inverses $G$ and $\tilde G_k$, $k=1, \, 2, \, 3$. We also define the following constants: 
\begin{eqnarray}
{\mathcal I}_k &=& \int_{-1/2}^{1/2} G_k(m_q e^{- \kappa' V(x_1)} , 1) \, dx_1, \quad k=1, \, 2, \, 3, \label{eq:I1} \\
{\mathcal I}_{12} &=& \int_{-1/2}^0 G_1(m_q e^{- \kappa' V(x_1)} , 1) \, dx_1 + \int_0^{1/2} G_2(m_q e^{- \kappa' V(x_1)} , 1) \, dx_1. \label{eq:I12}
\end{eqnarray}

We define the following class of travelling wave solutions of the NSH system  \eqref{eq:fl_rho_sm}-\eqref{eq:fl_al_sm}:  

\begin{definition}
Travelling-wave solutions of the NSH system  \eqref{eq:fl_rho_sm}-\eqref{eq:fl_al_sm} will be defined by 
\begin{equation} 
\rho = \rho(x_1), \quad u = u(x_1), \quad \alpha = \beta(x_1) + (x_2-\lambda t) Z, 
\label{eq:sol_strip_form}
\end{equation}
where the functions $\rho$, $u$ and $\beta$ are smooth (at least $C^1$), the $x_1$-component of the flow vanishes (in the sense of Eq. \eqref{eq:str_rho_2}), $\rho$ satisfies the normalization condition \eqref{eq:rho_normaliz} expressed here as 
\begin{equation}
\int_{-1/2}^{1/2} \rho(x_1) \, dx_1 = 1, 
\label{eq:rho_normaliz_strip}
\end{equation}
and $\beta(0) = 0$. 
\label{def:TW_strip}
\end{definition}

\begin{remark}
(i) The condition $Z \in 2 \pi {\mathbb Z}$ guarantees that $\alpha$ is $1$-periodic with respect to~$x_2$.

\smallskip
\noindent
(ii) Since $\alpha$ is defined up to an additive constant (only derivatives of $\alpha$ appear in the model), the condition $\beta(0) =0$ fixes this constant. 

\smallskip
\noindent
(iii) The condition that the $x_1$-component of the flow vanishes is expressed by Eq. \eqref{eq:str_rho_2}. In general, for solutions having structure given by \eqref{eq:sol_strip_form}, the $x_1$-component of the flow is constant. But if this constant in not zero, this implies that there are particle or fluid flows across the boundaries at $x_1 = \pm \frac{1}{2}$, which seems physically odd and which we discard here. 

\smallskip
\noindent
(iv) The travelling-wave character of these solutions appears in the phase equation only. The density and velocity are stationary fluid quantities. However, note that, at the level of the particle system (see \cite{degond2022topological}), these solutions are associated to actual motion of the particles.

\label{rem:strip}
\end{remark}

The existence of such travelling-wave solutions is stated in the following propositions. We first deal with the cases $\ell > 0$ and $\ell < 0$ and finally with the case $\ell = 0$ (i.e. stationary solutions).

\subsection{Case $\ell > 0$}
\label{subsec:ell>0} 

\begin{proposition}
We assume that $\ell > 0$. Suppose 
\begin{equation}
\frac{|bZ|}{c_1 {\mathcal I}} \leq 1. 
\label{eq:strip_exist}
\end{equation}
Then, there exists $\ell^* >0$ such that for all $\ell$ with $0 \leq \ell \leq \ell^*$, there exist travelling-wave solutions in the sense of Definition \ref{def:TW_strip}. For all such solutions, $\rho$ and $u_2$ are identical and given by 
\begin{equation}
\rho = G(C e^{- \kappa' V}, \ell), \qquad u_2 = - \frac{bZ}{c_1} (\rho + \ell),
\label{eq:express_rho}
\end{equation}
with $C$ uniquely determined by \eqref{eq:rho_normaliz_strip}. In particular, $\rho$ and $|u_2|$ are even, strictly decreasing on $[0,\frac{1}{2})$ and such that $\rho(x_1) \to 0$, $u_2(x_1)\to \lambda/c_1$ when $x_1 \to \frac{1}{2}$. Furthermore:
\begin{itemize}
\item[(i)] if $0 < \ell < \ell^*$, there exist exactly \textbf{two} such solutions associated with different $u_1$ and~$\beta$. The functions $u_1$ are even, have constant sign and are opposite to each other. The functions $\beta$ are odd and opposite to each other.
\item[(ii)] if $\ell = \ell^*$, there exist exactly \textbf{two} such solutions associated with different $u_1$ and~$\beta$. The functions $u_1$ are odd, change sign at  $x_1 = 0$ only and are opposite to each other. The functions $\beta$ are even, vanish at $x_1=0$ only and are opposite to each other. 
\end{itemize}
If $\ell > \ell^*$ or if \eqref{eq:strip_exist} is not satisfied, there is no such solution. 
\label{prop:strip}
\end{proposition}

\begin{remark}
From \eqref{eq:def_lam} and the second equation of \eqref{eq:express_rho}, we have
$$ u_2 = \frac{\lambda}{\ell \, c_1}(\rho + \ell), $$
which, since $\ell >0$ shows that $u_2$ has the same sign as $\lambda$, i.e., the fluid and the travelling wave move in the same direction along the strip. 
\label{rem:rel_sign_u2_lam_l>0}
\end{remark}

In Case (i), we denote by $u_1^+$ and $u_1^-$ the positive and negative solutions respectively and in Case (ii), we denote by $u_1^+$ the solution which is positive on $(-\frac{1}{2},0)$ and by $u_1^-$ the opposite one.

\begin{figure}[ht!]
\centering

$\mbox{}$ \vspace{-0.2cm} \hspace{0.cm}
\subfloat[$\ell < \ell^*$]{\includegraphics[trim={3.5cm 16cm 6.cm 4.5cm},clip,height= 4.8cm]{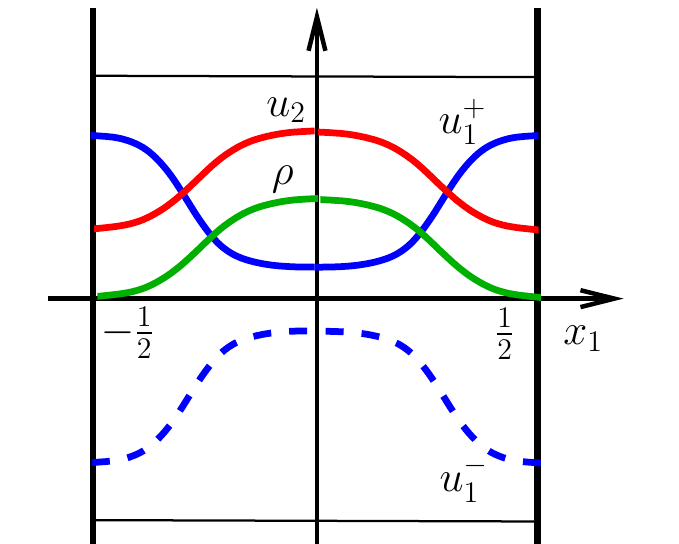}\label{subfig:strip_u1u2_l_leq_lV}} \hspace{0.9cm}
\subfloat[$\ell = \ell^*$]{\includegraphics[trim={3.5cm 16cm 6.cm 4.5cm},clip,height= 4.8cm]{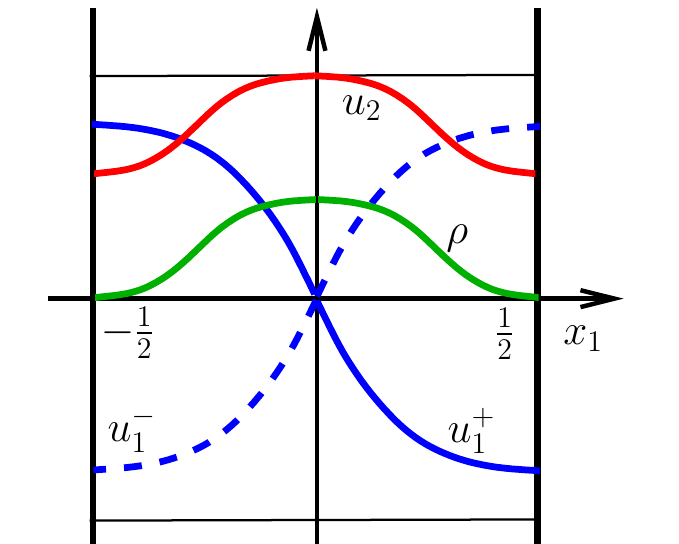}\label{subfig:strip_u1u2_l=lV}} 

$\mbox{}$ \vspace{-0.2cm} \hspace{0.1cm}
\subfloat[$\ell < \ell^*$]{\includegraphics[trim={3.5cm 14.cm 5.cm 4.5cm},clip,height= 5.8cm]{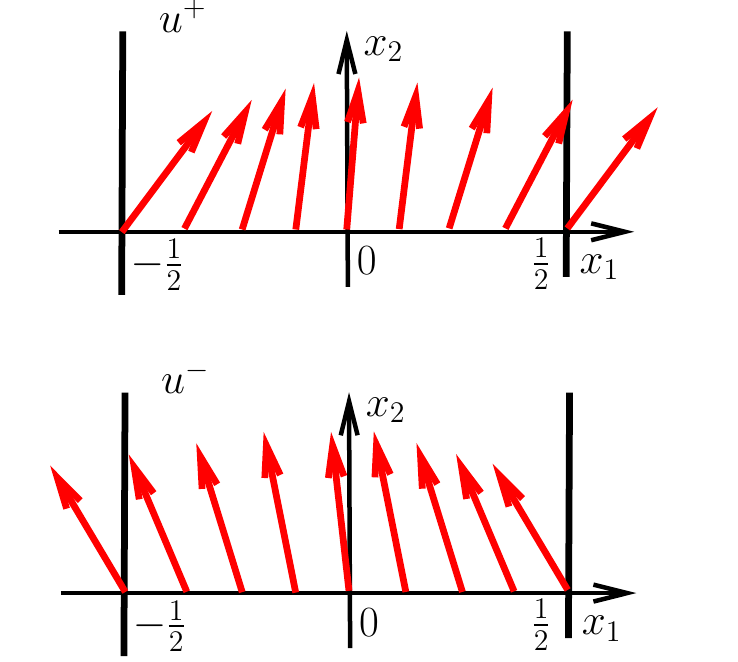}\label{subfig:strip_u_l_leq_lV}} \hspace{0.5cm}
\subfloat[$\ell = \ell^*$]{\includegraphics[trim={3.5cm 14.cm 5.cm 4.5cm},clip,height= 5.8cm]{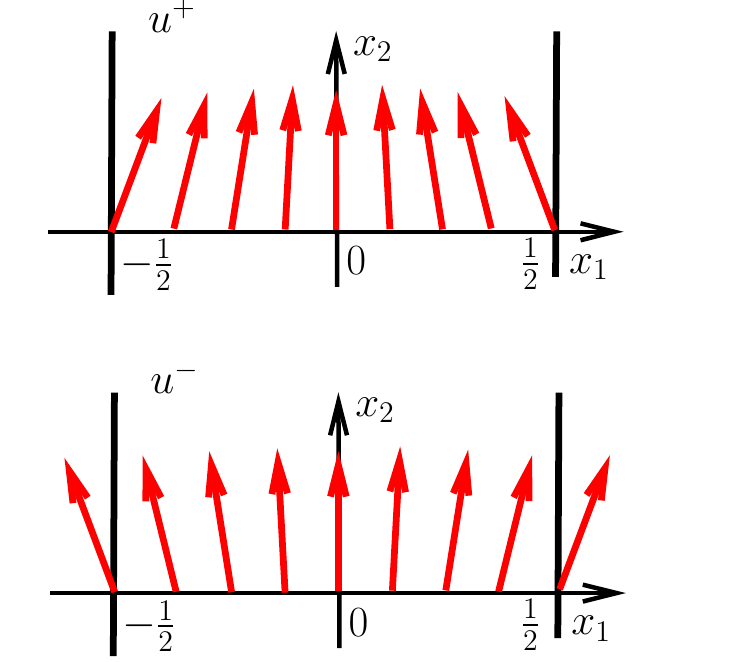}\label{subfig:strip_u_l=lV}}

$\mbox{}$ \hspace{0.5cm}
\subfloat[$\ell < \ell^*$]{\includegraphics[trim={3.5cm 16.cm 5.cm 4.5cm},clip,height=4.8cm]{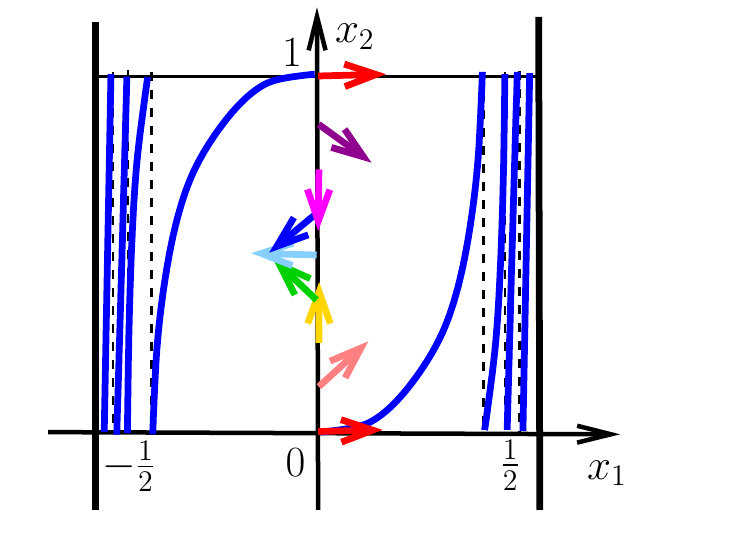}\label{subfig:strip_phase_l_leq_lV}} \hspace{0.35cm} 
\subfloat[$\ell = \ell^*$]{\includegraphics[trim={3.5cm 16.cm 5.cm 4.5cm},clip,height=4.8cm]{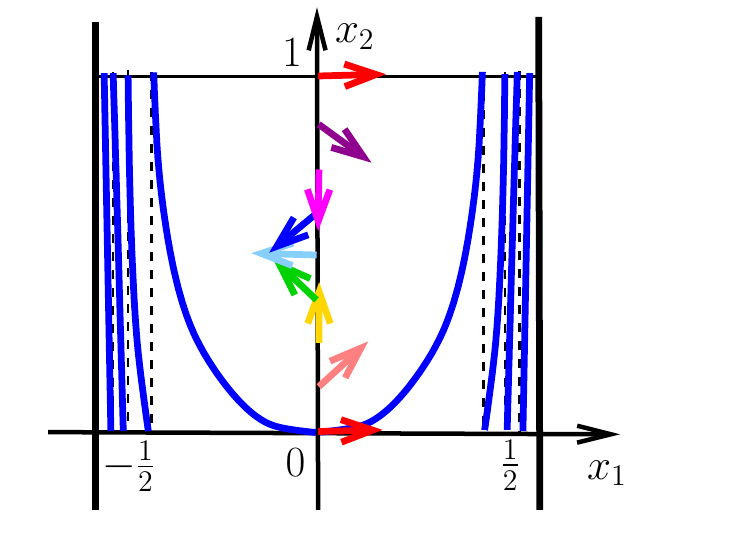}\label{subfig:strip_phase_l=lV}}
\caption{Structure of travelling-wave solutions for $\ell >0$ given by Prop. \ref{prop:strip}. (a), (c), (e): case $\ell < \ell^*$. (b), (d), (f): case $\ell = \ell^*$. (a), (b): sketches of $\rho$ (in green), $u_2$ (in red), $u_1^+$ (in blue) and $u_1^-$ (in blue, dashed line) as functions of $x_1$. Thin black horizontal line are drawn at ordinates $\pm 1$. (c), (d): velocities $u^+ = (u_1^+, u_2)$ (top) and $u^- = (u_1^-, u_2)$ (bottom) as functions of $x_1$. (e), (f):  vectors $e^{i \alpha}$ at $x_1=0$ and $t=0$ for $Z=2 \pi$ and $x_2 = k/8$, $k \in \{0, 1, \ldots, 8\}$ (see text for the color code) and isoline of $\alpha$ passing through the origin at time $t=0$ (in blue) in the domain~$\Omega = (-\frac{1}{2},\frac{1}{2}) \times [0,1)$.}
\label{fig:strip_ualpha}
\end{figure}

The proof of this proposition is given in Section \ref{sec_strip_proofs}. The structure of these solutions is shown in Fig. \ref{fig:strip_ualpha}. The density $\rho$ (in green) and the component $u_2$ of the velocity (in red) have the same shape, save for the fact that $\rho$ vanishes at the domain boundaries, while $u_2$ converges to a finite value.  The differences between the case $\ell < \ell^*$ (left figures) and the case $\ell = \ell^*$ (right figures) is striking. The maximum value of $u_2$ is less than~$1$ in Fig. \ref{subfig:strip_u1u2_l_leq_lV} and is equal to $1$ in Fig. \ref{subfig:strip_u1u2_l=lV}. As a consequence, the component $u_1$ of the velocity (in solid or dashed blue because there are two possible solutions) have completely different shapes. Indeed, while $u_1^\pm$ are even functions of $x_1$ in the case $\ell < \ell^*$ (Fig. \ref{subfig:strip_u1u2_l_leq_lV}), they are odd in the case $\ell = \ell^*$ (Fig. \ref{subfig:strip_u1u2_l=lV}). It results in the associated vector fields $u^\pm$ pointing towards the same horizontal direction throughout the domain in the case $\ell < \ell^*$ (Fig. \ref{subfig:strip_u_l_leq_lV}) while they point to opposite horizontal directions according to whether $x_1$ is positive or negative if $\ell = \ell^*$ (Fig. \ref{subfig:strip_u_l=lV}). It also results in different behavior of the isolines of the phase $\alpha$ in the two cases as shown in Figs \ref{subfig:strip_phase_l_leq_lV} and~\ref{subfig:strip_phase_l=lV}. In Figs \ref{subfig:strip_phase_l_leq_lV} and \ref{subfig:strip_phase_l=lV}, we have also drawn the phase vector $e^{i \alpha}$ at $x_1=0$ and $t=0$ for a discrete set of points $x_2 = k/8$, $k \in \{0, 1, \ldots, 8\}$ in the case $Z = 2 \pi$. The color code corresponds to the angle $\alpha$ with red, pink, yellow, green, light blue, blue, magenta, purple corresponding to $\alpha = 0, \, \frac{\pi}{4}, \frac{\pi}{2}, \frac{3 \pi}{4}, \pi, \frac{5\pi}{4}, \frac{3\pi}{2}, \frac{7\pi}{4}$ respectively. As expected, the index of the phase vector is $1$ as one moves one period along the $x_2$-axis, showing that these travelling-wave solutions are endowed with a non-trivial topology. Along a line $x_2= $ Constant, the behavior of the phase isolines indicates that the phase vector makes an infinite number of rotations as one approaches the boundary. Furthermore the direction of rotation does not change in the case $\ell < \ell^*$ while it reverses at $x_1=0$ in the case $\ell = \ell^*$.

\subsection{Case $\ell < 0$}
\label{subsec:ell<0}

\begin{proposition}
We assume that $\ell < 0$ and let $\tilde \ell = - \ell >0$. Then, there are four classes of solutions
\begin{itemize}
\item Class (a): There exists $\tilde \ell^*_a \in [0,1]$, with $\tilde \ell^*_a = 0$ if \eqref{eq:strip_exist} is satisfied and $\tilde \ell^*_a > 0$ otherwise, such that for all $\tilde \ell \in [\tilde \ell^*_a,1]$ (and $\tilde \ell \not = 0$), there exist travelling wave solutions in the sense of Definition \ref{def:TW_strip}. For all such solutions, $\rho$ and $u_2$ are identical and given by 
\begin{equation}
\rho = G_3(C e^{- \kappa' V}, \tilde \ell), \qquad u_2 = - \frac{bZ}{c_1} (\rho - \tilde \ell), 
\label{eq:express_rho_3}
\end{equation}
with $C$ uniquely determined by \eqref{eq:rho_normaliz_strip}. In particular, $\rho$ and $|u_2|$ are even, strictly decreasing on $[0,\frac{1}{2})$ and such that $\rho(x_1) \to \tilde \ell$, $u_2(x_1) \to 0$ when $x_1 \to \frac{1}{2}$. Furthermore:
\begin{itemize}
\item[(i)] If $0\leq\tilde \ell^*_a < \tilde \ell \leq 1$ or ($0 = \tilde \ell^*_a = \tilde \ell$ and ${\mathcal I}^{-1} < \frac{c_1}{|bZ|}$), there exist exactly \textbf{two} such solutions associated with different $u_1$ and $\beta$. The functions $u_1$ are even, have constant sign and are opposite to each other. The functions $\beta$ are odd and opposite to each other.
\item[(ii)] If $0 < \tilde \ell^*_a = \tilde \ell \leq 1$ or ($0 = \tilde \ell^*_a = \tilde \ell$ and ${\mathcal I}^{-1} = \frac{c_1}{|bZ|}$), there exist exactly \textbf{two} such solutions. The functions $u_1$ are odd, change sign at  $x_1 = 0$ only and are opposite to each other. The functions $\beta$ are even, vanish at $x_1=0$ only and are opposite to each other. 
\end{itemize}
If $\tilde \ell \not \in [\tilde \ell^*_a,1]$, there is no such solution.

\item Class (b): Assume that 
\begin{equation}
\frac{|bZ|}{c_1 {\mathcal I}_1} \leq 1. 
\label{eq:strip_exist_classb}
\end{equation}
Then, for any $\tilde \ell \in [{\mathcal I}_1^{-1}, \frac{c_1}{|bZ|}]$, there exist travelling wave solutions in the sense of Definition \ref{def:TW_strip}. For such solutions, $\rho$ and $u_2$ are identical and given by 
\begin{equation}
\rho = G_1(C e^{- \kappa' V}, \tilde \ell), \qquad u_2 = \frac{bZ}{c_1} (\tilde \ell - \rho), 
\label{eq:express_rho_1}
\end{equation}
with $C$ uniquely determined by \eqref{eq:rho_normaliz_strip}. In particular, $\rho$ and $u_2$ are even, $\rho$ is strictly decreasing and $u_2$, strictly increasing on $[0,\frac{1}{2})$ and verify $\rho(x_1) \to 0$, $u_2(x_1) \to \lambda/c_1$ when $x_1 \to \frac{1}{2}$. Furthermore, there exist exactly \textbf{two} such solutions associated with different $u_1$ and $\beta$. The functions $u_1$ are even, have constant sign and are opposite to each other. The functions $\beta$ are odd and opposite to each other.

\noindent
If $\tilde \ell \not \in [{\mathcal I}_1^{-1}, \frac{c_1}{|bZ|}]$ or if \eqref{eq:strip_exist_classb} is not satisfied, there is no such solution.

\item Class (c): There exists $\tilde \ell^*_c \in [1,{\mathcal I}_2^{-1}]$, with $\tilde \ell^*_c = {\mathcal I}_2^{-1}$ if the condition 
\begin{equation}
\frac{(1-q)|bZ|}{c_1 {\mathcal I}_2} \leq 1, 
\label{eq:cond_class_c}
\end{equation}
is satisfied and $\tilde \ell^*_c < {\mathcal I}_2^{-1}$ otherwise, such that for all $\tilde \ell \in [1,\tilde \ell^*_c]$, there exist travelling wave solutions in the sense of Definition \ref{def:TW_strip}. For such solutions, $\rho$ and $u_2$ are identical and given by 
\begin{equation}
\rho = G_2(C e^{- \kappa' V}, \tilde \ell), \qquad u_2 = \frac{bZ}{c_1} (\tilde \ell - \rho ), 
\label{eq:express_rho_2}
\end{equation}
with $C$ uniquely determined by \eqref{eq:rho_normaliz_strip}. In particular, $\rho$ and $|u_2|$ are even, $\rho$ is strictly increasing  and $|u_2|$, strictly decreasing on $[0,\frac{1}{2})$. They are such that $\rho(x_1) \to \tilde \ell$, $u_2(x_1) \to 0$ when $x_1 \to \frac{1}{2}$. Furthermore:
\begin{itemize}
\item[(i)] if $ 1 \leq \tilde \ell < \tilde \ell^*_c < {\mathcal I}_2^{-1}$ or ($\tilde \ell = \tilde \ell_c^* = {\mathcal I}_2^{-1}$ and $(1-q) {\mathcal I}_2^{-1} < \frac{c_1}{|bZ|}$), there exist exactly \textbf{two} such solutions associated with different $u_1$ and $\beta$. The functions $u_1$ are even, have constant sign and are opposite to each other. The functions $\beta$ are odd and opposite to each other.
\item[(ii)] if $1 \leq \tilde \ell = \tilde \ell^*_c < {\mathcal I}_2^{-1}$ or ($\tilde \ell = \tilde \ell^*_c={\mathcal I}_2^{-1}$ and $(1-q) {\mathcal I}_2^{-1} = \frac{c_1}{|bZ|}$), there exist exactly \textbf{two} such solutions. The functions $u_1$ are odd, change sign at  $x_1 = 0$ only and are opposite to each other.  The functions $\beta$ are even, vanish at $x_1=0$ only and are opposite to each other. 
\end{itemize}
If $\tilde \ell \not \in [1, \tilde \ell^*_c]$, there is no such solution.

\item Class (d): we assume that 
\begin{equation}
\frac{|bZ|}{c_1 {\mathcal I}_{12}} \leq 1 \quad \textrm{ and } \quad \tilde \ell =  {\mathcal I}_{12}^{-1}. 
\label{eq:cond_class_d}
\end{equation}
Then, there exist travelling-wave solutions in the sense of Definition \ref{def:TW_strip}. Additionally, $\rho$ and $u_2$ are given either by 
\begin{equation}
\rho(x_1) = \left\{ \begin{array}{lll} G_1  (M_{\tilde \ell} e^{-\kappa' V}, \tilde \ell) & \textrm{if} & x_1 \leq 0,  \\
 G_2  (M_{\tilde \ell} e^{-\kappa' V}, \tilde \ell) & \textrm{if} & x_1 \geq 0,  
\end{array} \right. \qquad u_2 = \frac{bZ}{c_1} (\tilde \ell - \rho), 
\label{eq:express_rho_4}
\end{equation}
or by the formula where $G_1$ and $G_2$ are exchanged. In particular, for the solution given by \eqref{eq:express_rho_4}, $\rho$ is strictly increasing on $(-1/2,1/2)$ and $u_2$ strictly decreasing, and are such that $\rho(x_1) \to 0$, $u_2(x_1) \to \lambda/c_1$ as $x_1 \to -1/2$ and $\rho(x_1) \to {\mathcal I}_{12}^{-1}$, $u_2(x_1) \to 0$ as $x_1 \to 1/2$. Furthermore, for $\rho$ and $u_2$ given by \eqref{eq:express_rho_4}, there are exactly two such solutions, corresponding to opposite functions $u_1$ and $\beta$. The functions $u_1$ have constant sign. When $\rho$ and $u_2$  are given by \eqref{eq:express_rho_4} where the roles of $G_1$ and $G_2$ are exchanged, there are also two solutions corresponding to opposite $u_1$  and $\beta$. If \eqref{eq:cond_class_d} is not satisfied, there is no such travelling-wave solution. 
\end{itemize}
\label{prop:strip_negative}
\end{proposition}

\begin{remark} Like in Remark \ref{rem:rel_sign_u2_lam_l>0}, we can write $u_2$ as follows: 
\begin{itemize} 
\item Class (a) of solutions: 
$$ u_2 = - \frac{\lambda}{\tilde \ell \, c_1}(\rho - \tilde \ell), $$
which, since $\tilde \ell >0$ shows that $u_2$ has opposite sign to $\lambda$. Hence, the fluid and the travelling wave move in opposite directions along the strip.
\item Classes (b), (c) and (d) of solutions: 
$$ u_2 = \frac{\lambda}{\tilde \ell \, c_1}(\tilde \ell - \rho). $$
In these cases, $u_2$ has the same sign as $\lambda$, which correspons to the fluid and the travelling wave moving in the same direction along the strip.
\end{itemize}
\label{rem:rel_sign_u2_lam_l<0}
\end{remark}

The proof of this proposition is given in Section \ref{sec_strip_proofs}. We denote by $u_1^+$ (resp. $u_1^-$) the positive (resp. negative) function $u_1$ in Classes (a)(i), (b), (c)(i), (d) and the function $u_1$ which is positive (resp. negative) on $(-1/2,0)$ in Classes (a)(ii) and (c)(ii). Finally, in Case (d), we denote by $\rho^+$ the increasing $\rho$ solution and by $\rho^-$the decreasing one. 

\begin{figure}[htbp]
\centering

$\mbox{}$ \vspace{-0.2cm} \hspace{0.cm}
\subfloat[Class (a), $0 \leq \tilde \ell^*_a < \tilde \ell < 1$]{\includegraphics[trim={3.5cm 16cm 6.cm 4.5cm},clip,height= 4.2cm]{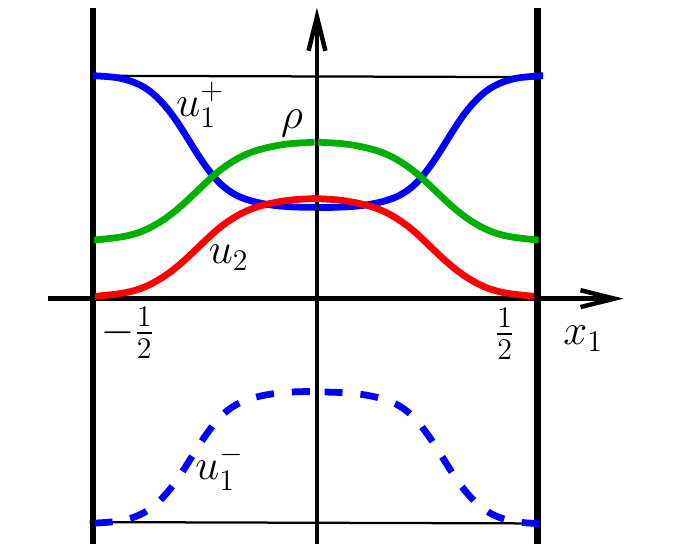}\label{subfig:Case(a)_l_geq_lV}} \hspace{0.9cm}
\subfloat[Class (a), $0 < \tilde \ell^*_a = \tilde \ell $]{\includegraphics[trim={3.5cm 16cm 6.cm 4.5cm},clip,height= 4.2cm]{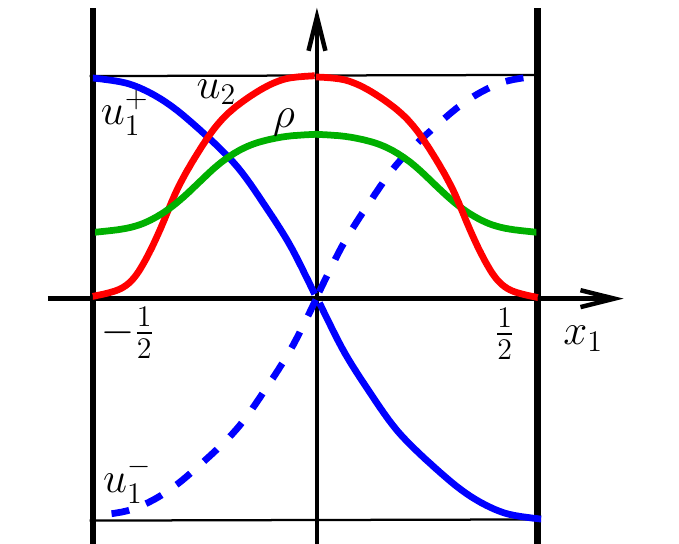}\label{subfig:Case(a)_l=lV}} 

$\mbox{}$ \vspace{-0.2cm} \hspace{0.cm}
\subfloat[Class (b), ${\mathcal I}_1^{-1} \leq \tilde \ell < \frac{c_1}{|bZ|}$]{\includegraphics[trim={3.5cm 16cm 6.cm 4.5cm},clip,height= 4.2cm]{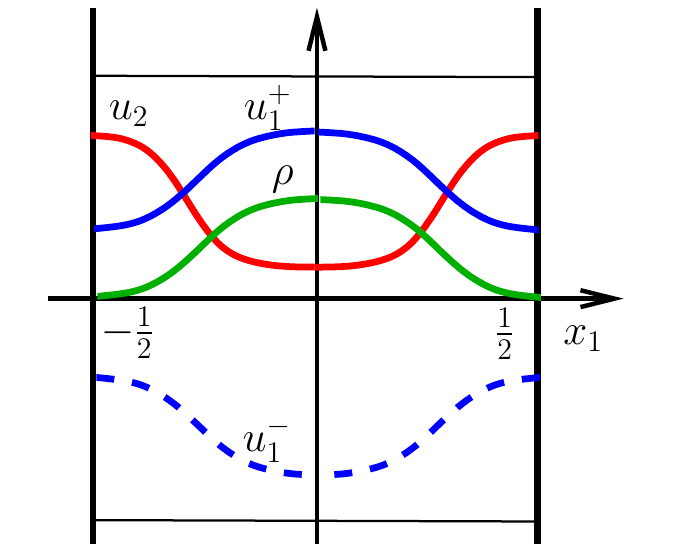}\label{subfig:Case(b)_l_geq_lV}} \hspace{0.9cm}
\subfloat[Class (b), $\tilde \ell = \frac{c_1}{|bZ|}$]{\includegraphics[trim={3.5cm 16cm 6.cm 4.5cm},clip,height= 4.2cm]{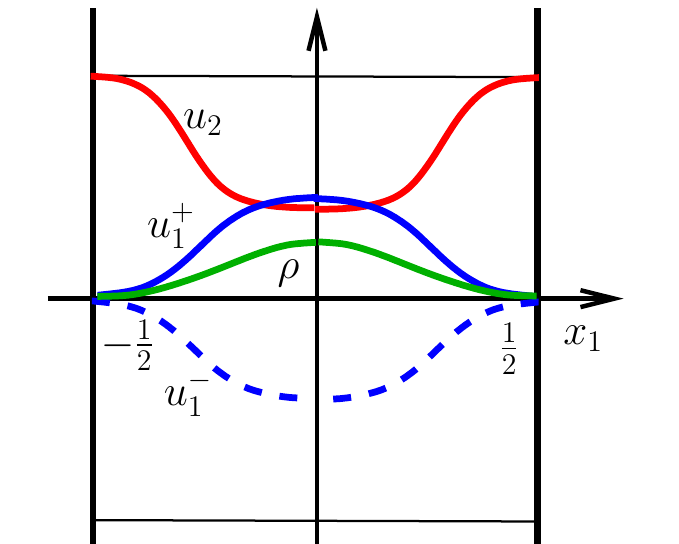}\label{subfig:Case(b)_l=lV}} 

$\mbox{}$ \vspace{-0.2cm} \hspace{0.cm}
\subfloat[Class (c), $1 \leq \tilde \ell< \tilde \ell^*_c$]{\includegraphics[trim={3.5cm 16cm 6.cm 4.5cm},clip,height= 4.2cm]{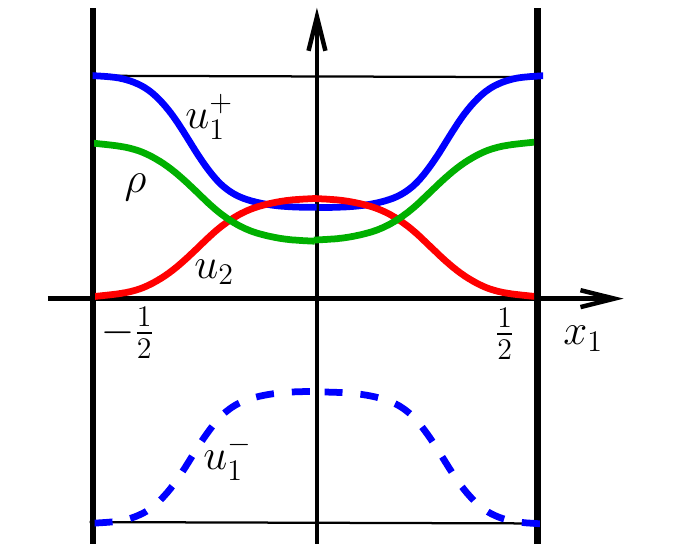}\label{subfig:Case(c)_l_leq_lV}} \hspace{0.9cm}
\subfloat[Class (c), $\tilde \ell = \tilde \ell^*_c$]{\includegraphics[trim={3.5cm 16cm 6.cm 4.5cm},clip,height= 4.2cm]{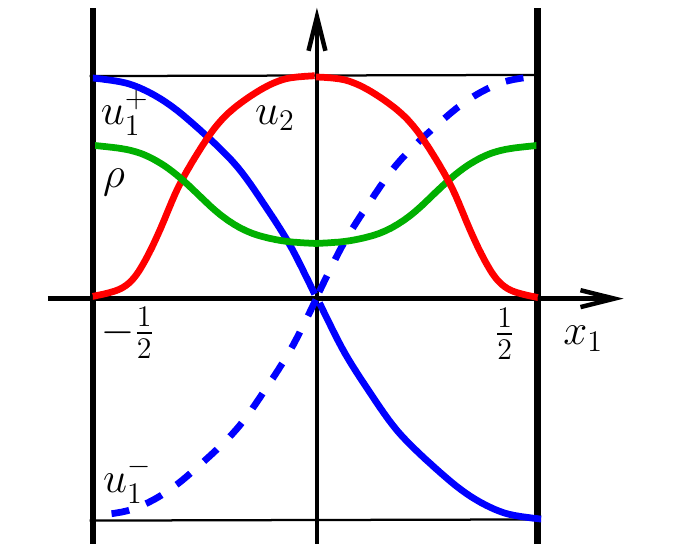}\label{subfig:Case(c)_l=lV}} 

$\mbox{}$ \vspace{-0.2cm} \hspace{0.cm}
\subfloat[Class (d), $\frac{|bZ|}{c_1 {\mathcal I}_{12}} < 1$]{\includegraphics[trim={3.5cm 16cm 6.cm 4.5cm},clip,height= 4.2cm]{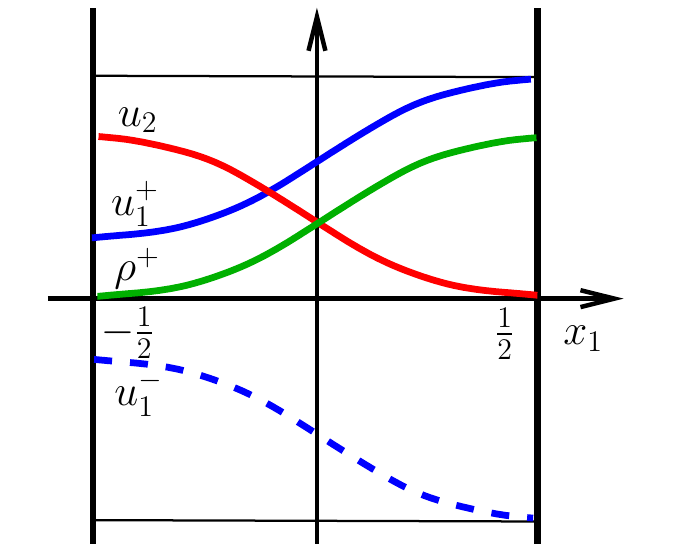}\label{subfig:Case(d)_l_geq_lV}} \hspace{0.9cm}
\subfloat[Class (d), $\frac{|bZ|}{c_1 {\mathcal I}_{12}} = 1$]{\includegraphics[trim={3.5cm 16cm 6.cm 4.5cm},clip,height= 4.2cm]{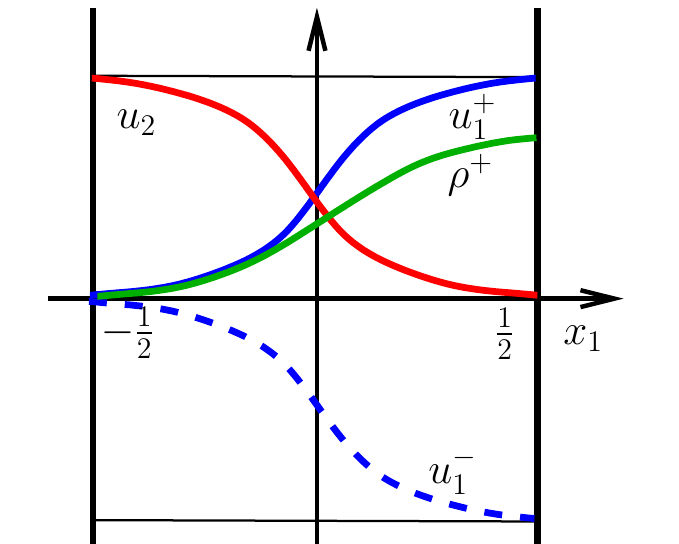}\label{subfig:Case(d)_l=lV}} 

\caption{Structure of travelling-wave solutions in the case $\tilde \ell = - \ell >0$  given by Prop.~\ref{prop:strip}. Sketches of $\rho$ (or $\rho^+$ for Class (d)) (in green), $u_2$ (in red), $u_1^+$ (in blue) and $u_1^-$ (in blue, dashed line) as functions of~$x_1$. Thin black horizontal lines at ordinates $\pm1$. }
\label{fig:case_l_negative}
\end{figure}

The solutions are depicted in Fig. \ref{fig:case_l_negative}. This figure highlights that Class (a) (respectively Class (c)) solutions have qualitatively different behaviors in the limiting cases $\tilde \ell = \tilde \ell^*_a$ (resp. $\tilde \ell = \tilde \ell^*_c$) and in the non-limiting ones $\tilde \ell > \tilde \ell^*_a$ (resp. $\tilde \ell < \tilde \ell^*_c$), as shown by $u_1^\pm$ changing sign in the former case (see Figs. \ref{subfig:Case(a)_l=lV} and \ref{subfig:Case(c)_l=lV}) and having constant sign in the latter (see Figs.~\ref{subfig:Case(a)_l_geq_lV} and \ref{subfig:Case(c)_l_leq_lV}). The situation is similar to what we observed in the case $\ell >0$ (see Fig. \ref{subfig:strip_u1u2_l=lV} versus Fig. \ref{subfig:strip_u1u2_l_leq_lV}). On the other hand, the limiting cases in Classes (b) and (d) (which correspond to $u_2$ reaching the value $1$ and are the cases $\tilde \ell = \frac{c_1}{|bZ|}$ in Class (b) and $\frac{|bZ|}{c_1 {\mathcal I}_{12}} = 1$ in Class~(d)) are not different from the non-limiting ones (respectively $\tilde \ell < \frac{c_1}{|bZ|}$ for Class~(b) and $\frac{|bZ|}{c_1 {\mathcal I}_{12}} < 1$ for Class (d)). Indeed, for these two classes, the functions $u_1^\pm$ change sign in the domain in neither the limiting case, nor the non-limiting one (compare Fig. \ref{subfig:Case(b)_l=lV} to Fig. \ref{subfig:Case(b)_l_geq_lV} for Class (b) and Fig. \ref{subfig:Case(d)_l=lV} to Fig. \ref{subfig:Case(d)_l_geq_lV} for Class (d)). So, Classes (b) and~(d) form homogeneous classes of solutions, by contrast to Classes (a) and (c).

\setcounter{equation}{0}
\section{Travelling-waves in annular geometry}
\label{sec:explicit_annular}

In this section, we still focus on dimension $n=2$ and let $(r,\vartheta)$ be the polar coordinates of the point $x \not = 0$, i.e. $x_1 = r \cos \vartheta$, $x_2 = r \sin \vartheta$. We denote by $(e_r,e_\vartheta)$ with $e_r = (\cos \vartheta, \sin \vartheta)^T$ and $e_\vartheta = (- \sin \vartheta, \cos \vartheta)^T$ the local polar basis and by $(u_r,u_\vartheta)$ the two coordinates of the self-propulsion velocity $u$ in the local polar basis. We recall that $u$ is a normalized vector, i.e. 
\begin{equation}
u_r^2 + u_\vartheta^2 = 1.  
\label{eq:str_const_cyl}
\end{equation}
We consider a spatial domain $\Omega$ defined in polar coordinates by $\Omega = \{ x \in {\mathbb R}^2 \, \, | \, \, r \in (R_0, R_1) \}$ where $0<R_0 < R_1 < +\infty$ are given. We let $V = V(r)$ be defined for $R_0 < r < R_1$, smooth, strictly convex as a function of $r$ and such that $V(r) \to + \infty$ when  $r \to R_0$ and $r \to R_1$. We assume that $\min_{(R_0,R_1)} V = 0$ and we denote by $R^*$ the point at which $V$ reaches its minimum, i.e. $V(R^*) = 0$. We suppose that $V''(R^*)>0$. We assume that the function $r \mapsto \frac{V'(r)}{r}$ is strictly increasing on $[R^*, R_1)$ (where primes denote derivatives with respect to $r$). Finally, we assume that 
\begin{equation}
2c_2 - c_1 >0. 
\label{eq:2c2-c1}
\end{equation}
This condition is satisfied if the particle system has small noise in the equation defining the  self-propulsion direction of the agents (see \cite{degond2022topological}). We have the following 

\begin{proposition}
Let $m \in {\mathbb Z} \setminus \{0\}$. We assume that
\begin{equation}
\ell =: - \frac{\lambda}{bm} \geq 0.  
\label{eq:annular_def_ell}
\end{equation}
We define 
\begin{equation}
q = \frac{\Theta}{c_1 - c_2 + \Theta}, \quad \kappa' = \frac{\kappa}{c_1 - c_2 + \Theta}, \quad {\mathcal I}_1 = 2 \pi \int_{R_0}^{R_1} e^{- \kappa' \, V} \, r \, dr. 
\label{eq:annul_calI}
\end{equation}
Then, if the condition 
\begin{equation}
\frac{|bm|}{c_1 {\mathcal I}_1 R_0} \leq 1, 
\label{eq:annul_cond}
\end{equation}
is satisfied, there exists $\ell^*$, such that for all $0 \leq \ell \leq \ell^*$, there exist travelling wave solutions of the NSH system \eqref{eq:fl_rho_sm}-\eqref{eq:fl_al_sm} of the form: 
\begin{equation} 
\rho = \rho(r), \quad u = u_r(r) e_r + u_\vartheta(r) e_\vartheta, \quad \alpha = \beta(r) + m (\vartheta -\lambda t), 
\label{eq:sol_annul_form}
\end{equation}
such that $\rho$, $u$ and $\beta$ are smooth (at least $C^1$), the radial flow is zero and $\rho$ satisfies the normalization condition~\eqref{eq:rho_normaliz} (over the domain $\Omega$). Additionally, $\rho$ has a unique maximum located at a point $r_\ell$, is strictly increasing on $(R_0,r_\ell)$, strictly decreasing on $(r_\ell,R_1)$ and such that $\rho(r) \to 0$ when $r \to R_0$ and $r \to R_1$, while $u_\vartheta$ has the sign of $\lambda$ and $|u_\vartheta| = \frac{|bm|}{c_1 r} (\rho + \ell r^2)$. Furthermore:
\begin{itemize}
\item if $\ell < \ell^*$, there exist exactly \textbf{two} solutions (up to an additive constant for $\beta$) which verify $\max_{r\in [R_0,R_1]}|u_\vartheta (r)| <1$. These two solutions have identical $u_\vartheta$ and $\rho$. The functions $u_r$ have constant sign and are opposite to each other. We denote by $u_r^+$ and $u_r^-$ the positive and negative ones respectively. Setting $\beta(R^*) = 0$, the functions $\beta$ change sign at $r=R^*$ and are opposite to each other. 
\item if $\ell = \ell^*$,  there are exactly \textbf{two} solutions (up to an additive constant for $\beta$) which verify $\max_{r\in [R_0,R_1]}|u_\vartheta (r)| =1$. These two solutions have identical $u_\vartheta$ and $\rho$. The functions $u_r$ have non-constant sign and are opposite to each other. We denote by $u_r^+$ the one which is positive near $r=R_0$ and by $u^-$ the opposite one. The functions $\beta$ are opposite to each other (provided the additive constant is the same for the two solutions). 
\end{itemize}
Furthermore, there exists $\bar \ell \geq \ell^*$ such that for all $\ell > \bar \ell$, there is no such solution. 
\label{prop:rotat_sym}
\end{proposition}

\begin{remark}
The condition that the radial flow is zero is given by Eq. \eqref{eq:zero_current}. In general, the radial flow is constant but we discard the cases where this constant is not zero. 
\end{remark}

The proof of Prop. \ref{prop:rotat_sym} can be found in Section \ref{sec_rotat_sym_proof}. It differs significantly from the strip case of Section \ref{sec_strip_proofs} and actually, the results are weaker in the present case. Indeed, in the strip case, explicit formulas are available, but they are lacking in the annulus case due to the presence of inertia forces. The proof of Prop. \ref{prop:rotat_sym} relies more heavily on geometric and analytic arguments. 

\begin{figure}[ht!]
\centering
\includegraphics[trim={3.5cm 15cm 5.cm 4cm},clip,width=6cm]{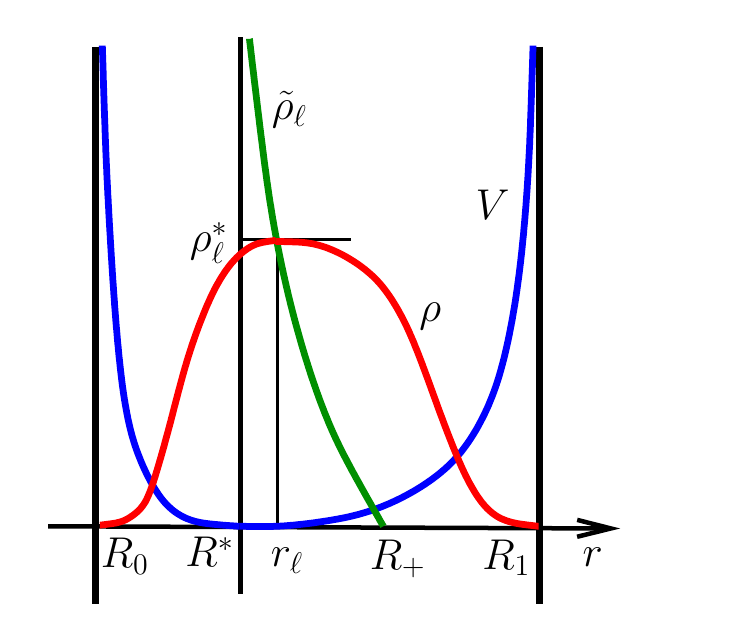}
\caption{External potential $V$ (in blue), density $\rho$ (in red) and function $\tilde \rho_\ell$ (in green) as functions of $r$ for a travelling-wave solution in an annulus given by Prop. \ref{prop:rotat_sym}. The function $\tilde \rho_\ell$ is the locus of the maximal points of $\rho$ when $\ell$ varies. The maximal point of $\rho$ has coordinates $(r_\ell, \rho^*_\ell)$ and $\rho^*_\ell= \tilde \rho_\ell(r_\ell)$. These definitions can be found in Section \ref{sec_rotat_sym_proof}.}
\label{fig:annulus_rhoV}
\end{figure}

\begin{figure}[ht!]
\centering

$\mbox{}$ \hspace{0.cm}
\subfloat[]{\includegraphics[trim={3.5cm 10cm 2.cm 4.5cm},clip,height= 6cm]{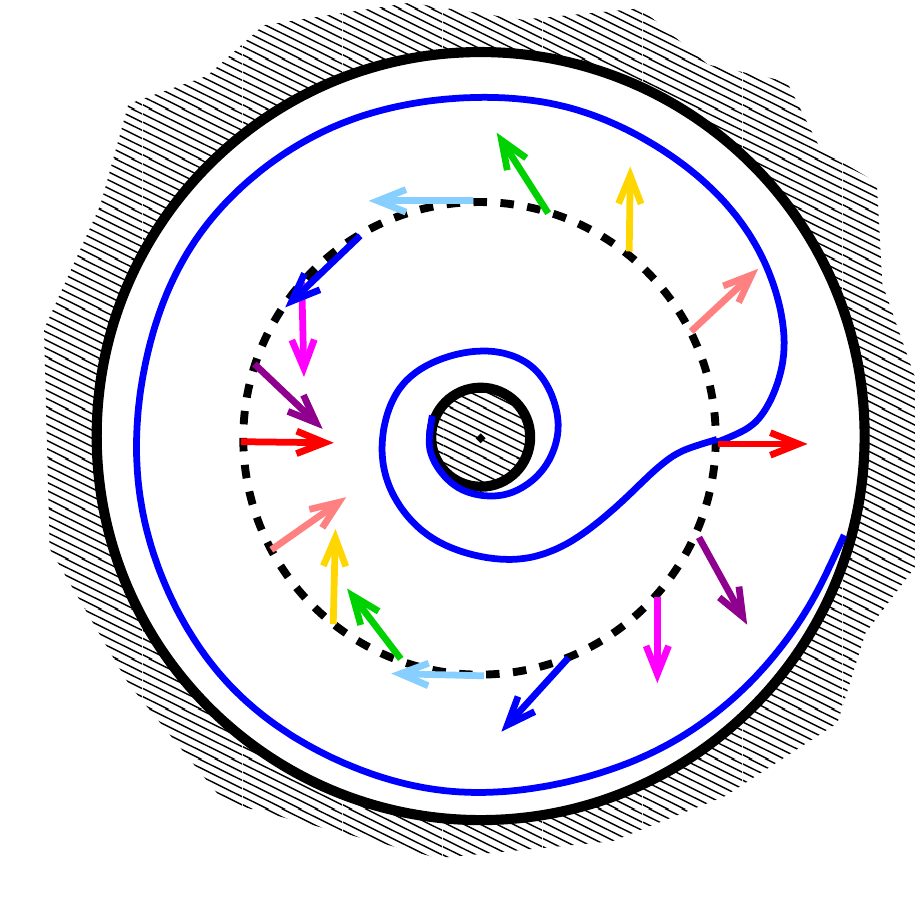}\label{subfig:annulus_phase_l_leq_lV}} \hspace{0.5cm}
\subfloat[]{\includegraphics[trim={3.5cm 10cm 2.cm 4.5cm},clip,height= 6cm]{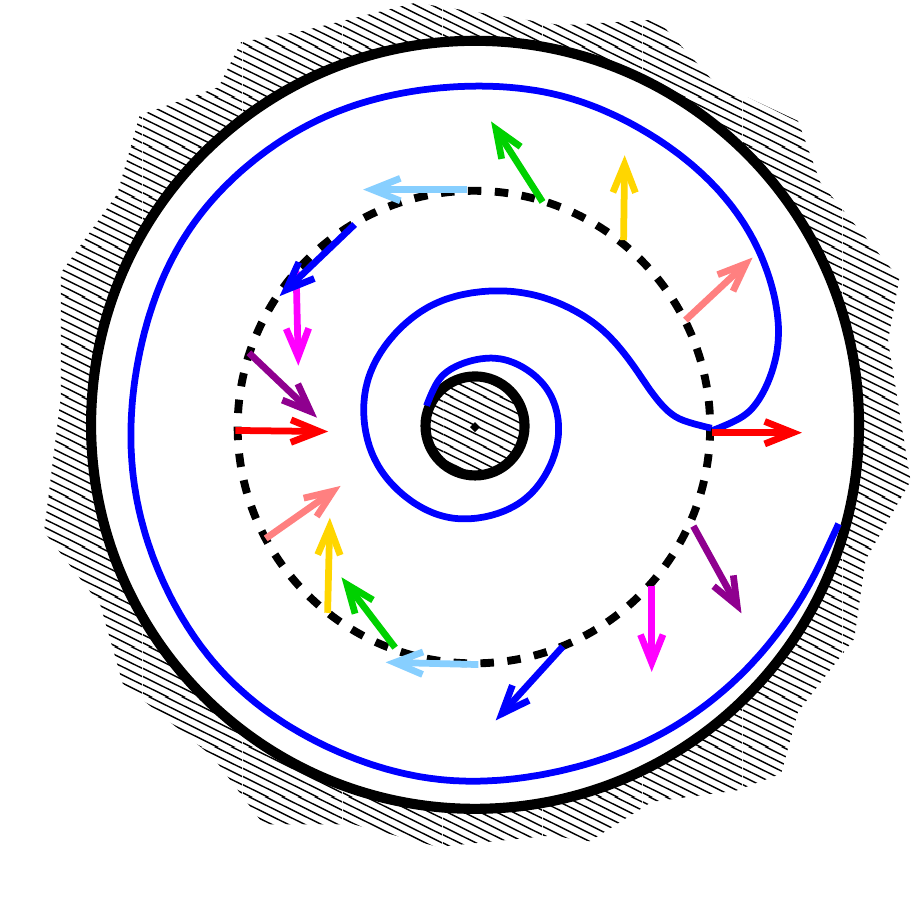}\label{subfig:annulus_phase_l=lV}} 

\caption{Structure of travelling-wave solutions in an annulus given by Prop. \ref{prop:rotat_sym}. (a):~case $\ell < \ell^*$. (b): case $\ell = \ell^*$. (a), (b):  vectors $e^{i \alpha}$ at $r=\hat r$ (where $\hat r$ is the maximal point of $u_\theta$) and $t=0$ for $m=2$, $\vartheta = k \pi/8$, $k \in \{0, 1, \ldots, 15\}$ and isolines of $\alpha$ passing through the point $(r,\vartheta) = (\hat r,0)$  at time $t=0$ (in blue) in the domain~$\Omega$. The color code corresponds to the angle $\alpha$ with red, pink, yellow, green, light blue, blue, magenta, purple corresponding to $\alpha = 0, \, \frac{\pi}{4}, \frac{\pi}{2}, \frac{3 \pi}{4}, \pi, \frac{5\pi}{4}, \frac{3\pi}{2}, \frac{7\pi}{4}$ respectively.}
\label{fig:annulus_phase}
\end{figure}

The structure of these solutions is given in Figs. \ref{fig:annulus_rhoV} and \ref{fig:annulus_phase}. By contrast to the case of a strip, the density $\rho$ (in red in Fig. \ref{fig:annulus_rhoV}) and the potential $V$ (in blue) do not reach their extrema at the same points. The maximum of $\rho$ lies on the green curve and is shifted to larger radii compared to the minimum of $V$. The components $u_r$ and $u_\vartheta$ of the self-propulsion velocity have roughly the same shape as in the case of a strip (see Figs. \ref{subfig:strip_u1u2_l_leq_lV} and \ref{subfig:strip_u1u2_l=lV}) but they are not symmetric (either odd or even) with respect to some intermediate point, and their end points at the values $r=R_0$ and $r=R_1$ are not equal or opposite. We have not been able to prove that their monotonicity is the same as in the strip case (with for instance $u_\vartheta$ increasing until reaching a maximum value at a certain point $\hat r$ and then, decreasing) but we we conjecture that it is indeed so. If this conjecture is correct, the profile of $u$ as a function of $r$ is roughly similar to that drawn in Figs. \ref{subfig:strip_u_l_leq_lV} and \ref{subfig:strip_u_l=lV}.

The phase $\alpha$ is schematically depicted in Fig. \ref{fig:annulus_phase}, assuming that $u_\vartheta$ has a unique maximum at $r = \hat r$. The phase vectors $e^{i \alpha}$ are drawn along the circle $r=\hat r$ (where $\hat r$ is the maximal point of $u_\theta$) at time $t=0$ for $m=2$ and polar angles $\vartheta = k \pi/8$, $k \in \{0, 1, \ldots, 15\}$ (see caption for the color code). The index of $e^{i \alpha}$  is equal to $2$ and shows that the solution drawn has non-trivial topology. The isolines of $\alpha$ passing through the point $(r,\vartheta) = (\hat r,0)$  at time $t=0$ are also drawn. These isolines wind around the origin as $r \to R_0$ and $r \to R_1$. However, while they roll up in the same direction in the case $\ell < \ell^*$ (Fig. \ref{subfig:annulus_phase_l_leq_lV}), the winding direction changes at $r=\hat r$ and they roll up in opposite directions in the case $\ell = \ell^*$ (Fig. \ref{subfig:annulus_phase_l=lV}). 

We cannot give an analog of Prop. \ref{prop:rotat_sym} for the case $\ell <0$. Indeed, the geometric arguments which led to Prop. \ref{prop:rotat_sym} cannot be developed in the same way due to the increased complexity of the phase portrait of the differential system to be solved. We refer to the end of Section \ref{sec_rotat_sym_proof} where these arguments are developed.

\setcounter{equation}{0}
\section{Proofs (strip geometry case)}
\label{sec_strip_proofs}

In this section, we give the proofs of Prop. \ref{prop:strip} and Prop. \ref{prop:strip_negative}. Inserting \eqref{eq:sol_strip_form} into the NSH ystem \eqref{eq:fl_rho_sm}-\eqref{eq:fl_al_sm} leads to 
\begin{eqnarray}
&&\hspace{-1cm}
\big[ \rho (c_1 u_1 + b \rho \beta') \big]' = 0, \label{eq:str_rho}\\
&&\hspace{-1cm}
(c_2 u_1 + b \rho \beta') u_1' + u_2^2 \big[ \Theta \log \rho + \kappa V \big]' = 0, \label{eq:str_u1}\\
&&\hspace{-1cm}
(c_2 u_1 + b \rho \beta') u_2' - u_1 u_2 \big[ \Theta \log \rho + \kappa V \big]' = 0, \label{eq:str_u2}\\
&&\hspace{-1cm}
- \lambda Z + (c_1 u_1 + b \rho \beta') \beta' + (c_1 u_2 + b \rho Z) Z = 0, \label{eq:str_bet} 
\end{eqnarray}
where primes denote derivatives with respect to $x_1$. The following condition 
\begin{equation}
c_1 u_1 + b \rho \beta' = 0, 
\label{eq:str_rho_2}
\end{equation}
implies \eqref{eq:str_rho}. Physically, this condition means that the component of the fluid velocity in the $x_1$-direction vanishes. From now on, we assume that \eqref{eq:str_rho_2} is satisfied. 

Now, multiplying \eqref{eq:str_u1} by $u_1$ and remarking that \eqref{eq:str_const} implies 
\begin{equation} 
u_1 \, u_1' = - u_2 \, u_2', 
\label{eq:udu}
\end{equation}
we get \eqref{eq:str_u2} provided that $u_2 \not = 0$. Thus, \eqref{eq:str_u1} and \eqref{eq:str_const} imply \eqref{eq:str_u2} wherever $u_2 \not = 0$. Using \eqref{eq:str_rho_2} and \eqref{eq:udu} again together with \eqref{eq:def_calI}, Eq. \ref{eq:str_u1} leads to 
$$
(1-q) u_2' + u_2 \big[ q \log \rho + \kappa' V \big]' = 0, 
$$
wherever $u_2 \not = 0$. Thus, there exists a constant $C > 0$ such that 
\begin{equation}
|u_2|^{1-q} \, \rho^q = C e^{- \kappa' V}. 
\label{eq:str_u1_2}
\end{equation}
Inserting \eqref{eq:str_rho_2} into \eqref{eq:str_bet} and using that $Z \not = 0$, we get 
\begin{equation}
u_2 = - \frac{bZ}{c_1} (\rho + \ell), 
\label{eq:str_bet_2}
\end{equation}
with $\ell$ given by \eqref{eq:def_lam}. Inserting \eqref{eq:str_bet_2} into \eqref{eq:str_u1_2}, we find that there exists another positive constant, still denoted by $C$, such that $\rho$ satisfies 
\begin{equation}
F(\rho,\ell) = C e^{- \kappa' V}, 
\label{eq:Fl(rho)}
\end{equation}
with $F$ given by \eqref{eq:definition_of_F}. Now, we discuss the sign of $\ell$.

\medskip
\paragraph{Case $\ell \geq 0$: proof of Prop. \ref{prop:strip}.}

In the case $\ell \geq 0$, the solution $\rho$ of \eqref{eq:Fl(rho)} is given by 
\begin{equation}
\rho = G \big( C e^{- \kappa' V}, \ell \big) =: \rho_{C, \ell}, 
\label{eq:str_rho_3}
\end{equation}
where $G$ is introduced at the beginning of Section \ref{subsec:setting}. We now show that the constant $C$ is uniquely determined by the normalization condition \eqref{eq:rho_normaliz_strip}. We define $I$: $[0,\infty)^2 \to [0,\infty)$, $(C,\ell) \mapsto I(C,\ell)$ by 
$$ I(C,\ell) = \int_{-1/2}^{1/2} \rho_{C, \ell} (x_1) \, d x_1 = \int_{-1/2}^{1/2} G \big( C e^{- \kappa' V(x_1)}, \ell \big) \, d x_1 . $$
For any $\ell \in [0,\infty)$, we have $F(0,\ell) = 0$, so, $G(0,\ell) = 0$. Consequently, $I(0,\ell) = 0$. Since $G(\cdot,\ell)$ is nonnegative and increasing, the family of functions $\rho_{C, \ell}$ is nonnegative and  increasing with respect to $C$ and so is $I(C,\ell)$. From $F(\rho,\ell) \sim \rho$ when $\rho \to +\infty$, we get $G(y,\ell) \sim y$ when $y \to +\infty$. So, $\rho_{C, \ell} \to +\infty$ as $C \to +\infty$ pointwise with respect to $x_1$. By the monotone convergence theorem, $I(C,\ell) \to +\infty$ when $C \to +\infty$. Thus, $I(\cdot,\ell)$ is increasing from $[0,\infty)$ onto $[0,\infty)$. Furthermore, by the dominated convergence theorem, $I$ belongs to $C^0\big([0,\infty)^2 \big) \cap C^\infty\big((0,\infty)^2 \big)$. Therefore, there exists a unique $C_\ell >0$ such that $I (C_\ell,\ell) = 1$ and, by the implicit function theorem, the map $[0,\infty) \to [0,\infty)$, $\ell \mapsto C_\ell$ belongs to $C^0\big([0,\infty) \big) \cap C^\infty\big((0,\infty) \big)$. We now denote by $\rho_\ell = \rho_{C_\ell,  \ell}$. 

We now return to \eqref{eq:str_bet_2} with $\rho = \rho_\ell$ and look for the conditions on $\ell$ such that $u_2$ satisfies the constraint $|u_2| \leq 1$. We note that, up to the factor $-\frac{bZ}{c_1}$, $u_2$ is just a translate of $\rho$ by the quantity $\ell$, so that $|u_2|$ has the same variations as $\rho$. From the assumptions on~$V$, the function $\rho_\ell$ is even and decreasing on $[0,1/2)$ from $\rho_\ell(0) = G(C_\ell, \ell)$ to $\rho_\ell(1/2)=0$. Thus, we have 
\begin{equation}
|u_2|(x_1) = \frac{|bZ|}{c_1} (\rho_\ell(x_1) + \ell) \leq \frac{|bZ|}{c_1} \big( G(C_\ell,\ell) + \ell \big) = |u_2|(0), \, \forall x_1 \in (-1/2,1/2). 
\label{eq:cond_u2}
\end{equation}
We denote by $H(\ell) = G(C_\ell,\ell) + \ell$. We show that $H$ is increasing. Indeed, 
$$ H'(\ell) = \frac{\partial G}{\partial \ell} (C_\ell,\ell) + \frac{\partial G}{\partial y} (C_\ell,\ell) \frac{d C_\ell}{d \ell} + 1.  $$
Since $G$ is increasing with respect to $y$, we have $\frac{\partial G}{\partial y} (C_\ell,\ell) \geq 0$. We also note that $F$ is increasing with respect to $\ell$. Since $F$ is also increasing with respect to $\rho$, it follows that $G$ is decreasing with respect to $\ell$. Then,  $I$ is also decreasing with respect to $\ell$ and since $I$ is increasing with respect to $C$, we deduce that $C_\ell$ is increasing with respect to $\ell$. Therefore, $\frac{d C_\ell}{d \ell} \geq 0$ and we finally get 
$$ H'(\ell) \geq \frac{\partial G}{\partial \ell} (C_\ell,\ell) + 1.  $$
Then, we compute
$$
\frac{\partial G}{\partial \ell} (y,\ell) = - \, \frac{\displaystyle \frac{\partial F}{\partial \ell} \big(G(y,\ell), \ell \big)}{\displaystyle \frac{\partial F}{\partial \rho} \big(G(y,\ell) ,\ell \big)} = - \frac{\displaystyle (1-q) G(y,\ell)}{\displaystyle G(y,\ell) + q \ell} \geq -1 + q . 
$$
Thus, $H'(\ell) \geq q >0$ as announced. 

Now, $H(0) = G(C_0,0)$. We have $F(\rho,0) = \rho$, so, $G (y,0) = y$. Then, $I(C,0) = C {\mathcal I}$, with ${\mathcal I}$ given by \eqref{eq:def_calI}. So, $H(0) = C_0 = {\mathcal I}^{-1}$. Furthermore, from $H'(\ell) \geq q >0$, we deduce that $H(\ell) \to + \infty$ as $\ell \to + \infty$. So, $H$ is continuous and increasing from $[0,\infty)$ onto $[{\mathcal I}^{-1} , \infty)$, and thus a bijection between these two sets. From \eqref{eq:cond_u2}, the condition that $|u_2| \leq 1$ is equivalent to the condition on $\ell$ that $H(\ell) \leq \frac{c_1}{|bZ|} $. This conditions can be only fulfilled if $\frac{c_1}{|bZ|} \geq {\mathcal I}^{-1}$. Hence, if \eqref{eq:strip_exist} is not satisfied, there is no such solution. If~\eqref{eq:strip_exist} is satisfied, the range of admissible $\ell$ is $[0, \ell^*]$ with $\ell^* = H^{-1}(\frac{c_1}{|bZ|} )$, and if $\ell > \ell^*$, there is no solution. 

We now find $u_1$. By \eqref{eq:str_const}, we get $u_1 = \sigma (1-u_2^2)^{1/2}$ where $\sigma \in \{-1,+1\}$ may depend on $x_1$. To preserve smoothness of $u_1$, $\sigma$ can only change when $|u_2| = 1$. 
\begin{itemize}
\item For $\ell < \ell^*$, we have 
$$ |u_2(x_1)| \leq |u_2(0)| = \frac{|bZ|}{c_1} H(\ell) <  \frac{|bZ|}{c_1} H(\ell^*)=1. $$
Thus, $|u_2|<1$ and $\sigma$ must be constant on $(-1/2,1/2)$. Thus, there are only two solution for $u_1$, denoted by $u_1^{\pm}$ corresponding to the two opposite choices of $\sigma$. 
\item For $\ell = \ell^*$, we have $|u_2(0)|=1$ and $|u_2(x_1)|<1$ for all $x_1 \not = 0$. Thus, $\sigma$ can change at $x_1 = 0$ and only there. In fact, we show that this sign must change to preserve smoothness of $u_1$. We first show that $u_2''(0) \not = 0$. Indeed, since $u_2$ is up to a constant and a multiplicative factor, equal to $\rho$, we show that $\rho''(0) \not = 0$. Indeed, differentiating \eqref{eq:Fl(rho)} twice with respect to $x_1$ at $x_1=0$ and using that $\rho'(0) = 0$ and $V'(0) = 0$ (because both $\rho$ and $V$ are even), $V(0)=0$ and $V''(0) \not = 0$ (by assumption on $V$), we get
$$ \frac{\partial F}{\partial \rho} (\rho(0), \ell) \rho''(0) = -C \kappa' V''(0) \not = 0, $$
which implies that $\rho''(0) \not = 0$ because  $\frac{\partial F}{\partial \rho} (\rho(0), \ell)$ is positive and finite. Thus, we have the following Taylor expansion of $|u_2|$ in the neighborhood of $x_1=0$: $|u_2|(x_1) = 1 - D x_1^2 + o(x_1^2)$ with $D  = |u_2''(0)|/2 >0$. So, $|u_1|(x_1) = (1-u_2^2)^{1/2} = \sqrt{D} |x_1| + o(|x_1|)$. If we take $u_1= \sqrt{D} |x_1| + o(|x_1|)$ or $u_1= - \sqrt{D} |x_1| + o(|x_1|)$ we lose the smoothness property (the solution is not $C^1$). Thus, the only possibilities are to take $u_1 = \sqrt{D} x_1 + o(|x_1|)$ or $u_1 = - \sqrt{D} x_1 + o(|x_1|)$, i.e. $u_1$ must change sign at $x_1 = 0$. Thus, we find two solutions for $u_1$: 
\begin{equation}
u_1^\searrow = \left\{ \begin{array}{ccc} (1-u_2^2)^{1/2} & \textrm{if} & x_1 <0 \\
- (1-u_2^2)^{1/2} & \textrm{if} & x_1 >0 \end{array} \right. , \quad u_1^\nearrow = -  u_1^\searrow. 
\end{equation}
\end{itemize}

Once $u_1$ is selected, $\beta$ is uniquely determined by \eqref{eq:str_rho_2}, i.e. by integrating
\begin{equation} 
\beta' = - \frac{c_1}{b} \frac{u_1}{\rho}, 
\label{eq:integration_beta'}
\end{equation}
with respect to $x_1$ together with the condition $\beta(0)=0$. This completes the construction of the travelling-wave solutions  \eqref{eq:sol_strip_form} of System \eqref{eq:fl_rho_sm}-\eqref{eq:fl_al_sm} and ends the proof of Prop. \ref{prop:strip}.

\paragraph{Case $\ell < 0$: proof of Prop. \ref{prop:strip_negative}.}

Suppose now $\ell < 0$ and introduce $\tilde \ell = - \ell >0$. Then, for $\rho$ to solve \eqref{eq:Fl(rho)} and being smooth, we have the following possibilities: either
\begin{equation}
\rho = G_k(C e^{-\kappa' V}, \tilde \ell) =: \rho^k_{C,\tilde \ell}, 
\label{eq:rho_ell_negative}
\end{equation}
with $C \leq M_{\tilde \ell}$ if $k=1, \, 2$ and $C \in (0,\infty)$ for $k=3$; or 
\begin{equation}
\rho(x_1) = \left\{ \begin{array}{lll} G_1  (M_{\tilde \ell} e^{-\kappa' V}, \tilde \ell) & \textrm{if} & x_1 \leq 0 \\
 G_2  (M_{\tilde \ell} e^{-\kappa' V}, \tilde \ell) & \textrm{if} & x_1 \geq 0 
\end{array} \right\} =: \rho^4_{\tilde \ell}(x_1),
\label{eq:rho_4}
\end{equation}
or 
\begin{equation}
\rho^5_{\tilde \ell}(x_1) = \rho^4_{\tilde \ell}(- x_1).
\label{eq:rho_5}
\end{equation}
We recall the definition \eqref{eq:def_Mtill} of $M_{\tilde \ell}$. We define Classes (a), (b) and (c) of solutions as corresponding to $\rho^3$, $\rho^1$ and $\rho^2$ respectively, while Class (d) corresponds to both $\rho^4$ and $\rho^5$. 

\medskip
\textbf{Class (a):} We first look at the conditions under which there exists a unique $C = C_{\tilde \ell} \in (0,\infty)$ such that $\rho^3_{C,\tilde \ell}$ satisfies the normalization condition \eqref{eq:rho_normaliz_strip}. We proceed like in the case $\ell > 0$ and define
\begin{equation} 
I_3 (C, \tilde \ell) = \int_{-1/2}^{1/2} \rho^3_{C,\tilde \ell}(x_1) \, dx_1 = \int_{-1/2}^{1/2} G_3(C e^{-\kappa' V(x_1)}, \tilde \ell) \, dx_1. 
\label{eq:def_I3}
\end{equation}
We note that
\begin{equation}
y \mapsto G_3( y , \tilde \ell) \textrm{ is increasing on } [0,\infty), \, \,  \lim_{y \to +\infty} G_3( y , \tilde \ell) = + \infty, \, \, G_3( 0 , \tilde \ell) = \tilde \ell. \label{eq:G3_limits}
\end{equation} 
So, $I_3( \cdot , \tilde \ell)$ is increasing and 
\begin{equation} 
\lim_{C \to +\infty} I_3(C, \tilde \ell)  = + \infty, \qquad  I_3(0, \tilde \ell)  = \tilde \ell. 
\label{eq:prop_C_ltil}
\end{equation}
So, there exists $C \in (0,\infty)$ such that $I_3 (C, \tilde \ell) = 1$ if and only if $\tilde \ell \leq 1$ and in this case, such $C$ is unique, denoted by $C_{\tilde \ell}$. We let $\rho_{\tilde \ell} = \rho^3_{C_{\tilde \ell},\tilde \ell}$. 

We note that $\rho_{\tilde \ell} \geq \tilde \ell$. Then, by \eqref{eq:str_bet_2}, we have the second equation of \eqref{eq:express_rho_3} and $u_2$ has the sign of $- bZ = - \lambda/\tilde \ell$ which is opposite to the sign of $\lambda$. Furthermore, by \eqref{eq:rho_ell_negative} (with $k=3$) and \eqref{eq:G3_limits}, $\rho_{\tilde \ell}$ is even, decreasing on $[0,1/2)$ and such that $\lim_{x_1 \to 1/2} \rho_{\tilde \ell}(x_1) = \tilde \ell$. So $\rho_{\tilde \ell}$ is maximal at $x_1 = 0$. Thus, $|u_2(x_1)| = \frac{|bZ|}{c_1} (\rho_{\tilde \ell} - \tilde \ell)$ is maximal at $x_1 = 0$ and $\lim_{x_1 \to 1/2} u_2(x_1) = 0$. Thus, we have 
$$ \max_{[-1/2,1/2]} |u_2| = |u_2(0)| = \frac{|bZ|}{c_1} H_3(\tilde \ell) , \qquad H_3(\tilde \ell) = G_3(C_{\tilde \ell}, \tilde \ell) - \tilde \ell, $$
where $H_3$ is defined (like $C_{\tilde \ell}$) for $\tilde \ell \in [0,1]$. We examine under which conditions  
\begin{equation}
\max_{[-1/2,1/2]} |u_2| = \frac{|bZ|}{c_1} H_3(\tilde \ell) \leq 1, 
\label{eq:admissibility}
\end{equation}
which is the necessary and sufficient condition for $u_2$ to be an admissible solution. 

We first show that $H_3$ is decreasing on $[0,1]$. We have 
$$ H_3'(\tilde \ell) = \frac{\partial G_3}{\partial \tilde \ell} (C_{\tilde \ell}, \tilde \ell) + \frac{\partial G_3}{\partial y} (C_{\tilde \ell},{\tilde \ell}) \frac{d C_{\tilde \ell}}{d \tilde \ell} - 1.  $$
By contrast to the case $\ell >0$, $\tilde F$ is now decreasing with respect to $\tilde \ell$. It results that $C_{\tilde \ell}$ is also decreasing with respect to $\tilde \ell$. Thus, by a similar computation as in the case $\ell >0$, we get 
\begin{equation}
 H_3'(\tilde \ell) \leq \frac{\partial G_3}{\partial \tilde \ell} (C_{\tilde \ell}, \tilde \ell) - 1 = 
\frac{q (\tilde \ell - \rho)}{\rho - q \tilde \ell} \leq 0,  
\label{eq:prop_H3}
\end{equation}
with $\rho = G_3 (C_{\tilde \ell}, \tilde \ell) \geq \tilde \ell > q \tilde \ell$. Furthermore, equality in \eqref{eq:prop_H3} requires $\rho = \tilde \ell$, i.e. $G_3 (C_{\tilde \ell}, \tilde \ell) = \tilde \ell$, i.e., $C_{\tilde \ell} = 0$, which only occurs for $\tilde \ell = 1$ (see below). Thus, $H_3$ is decreasing on $[0,1]$. 

Now, when $\tilde \ell = 0$, like in the case $\ell >0$, we have $H_3(0) = C_0 = {\mathcal I}^{-1}$. When $\tilde \ell = 1$, we have $C_1 = 0$ by virtue of the second equation \eqref{eq:prop_C_ltil} and the fact that, by the definition of $C_1$,  $I_3(C_1,1)=1$. Thus, $G_3(C_1,1) = G_3(0,1) = 1$ and consequently, $H_3(1) = G_3(C_1,1) - 1 = 0$. Thus $H_3$: $[0,1] \to [0,{\mathcal I}^{-1}]$ is continuous, decreasing and onto. But $u_2$ satisfies the admissibility condition \eqref{eq:admissibility} if and only if $H_3(\tilde \ell) \leq \frac{c_1}{|bZ|}$. So, there exists $\tilde \ell \in [0,1]$ satisfying this condition if and only if 
\begin{itemize}
\item either ${\mathcal I}^{-1} \leq \frac{c_1}{|bZ|}$, i.e. condition \eqref{eq:strip_exist} is satisfied. In this case, any value of $\tilde \ell \in [0,1]$ leads to an admissible solution; in this case, we let $\tilde \ell^*_a = 0$, 
\item or if  ${\mathcal I}^{-1} > \frac{c_1}{|bZ|}$, $H_3(\tilde \ell) \in [0, \frac{c_1}{|bZ|}]$, i.e. $\tilde \ell \in [\tilde \ell^*_a, 1]$ with $\tilde \ell^*_a >0$ the unique solution of $H_3(\tilde \ell) = \frac{c_1}{|bZ|}$. 
\end{itemize}

Now, we turn towards defining $u_1$. The situation is similar to the case $\ell >0$. 
\begin{itemize}
\item If $0\leq\tilde \ell^*_a < \tilde \ell \leq 1$ or ($0 = \tilde \ell^*_a = \tilde \ell$ and ${\mathcal I}^{-1} < \frac{c_1}{|bZ|}$), then $\max_{[-1/2,1/2]} |u_2| <1$. Thus $u_2$ does not reach the value $1$, so, $u_1$ cannot change sign. In this case, there are two opposite solutions, one positive, one negative which are both even functions. 
\item If $0 < \tilde \ell^*_a = \tilde \ell \leq 1$ or ($0 = \tilde \ell^*_a = \tilde \ell$ and ${\mathcal I}^{-1} = \frac{c_1}{|bZ|}$), then $\max_{[-1/2,1/2]} |u_2| =1$. Thus $u_2$ reaches the value $1$ at $x_1=0$. Then, the smoothness requirement obliges $u_1$ to change sign. Consequently, there are still two opposite solutions, but they are odd and change sign only at $x_1 = 0$. 
\end{itemize}
The properties of $\beta$ follow from integrating \eqref{eq:integration_beta'} with respect to $x_1$ with initial condition $\beta(0) = 0$. This completes the determination of Class (a) of solutions.

\medskip
\textbf{Class (b):} The definition of $\rho^1_{C,\tilde \ell}$ requires $C \leq M_{\tilde \ell}$ with $M_{\tilde \ell}$ given by \eqref{eq:def_Mtill}. We seek the conditions on $C$ such that the normalization condition \eqref{eq:rho_normaliz_strip} is satisfied. We define $I_1(C, \tilde \ell)$ by \eqref{eq:def_I3} with $G_3$ substituted by $G_1$. We note the following properties of $G_1$: 
\begin{equation}
y \mapsto G_1( y , \tilde \ell) \textrm{ is increasing on } [0,M_{\tilde \ell}], \, \,  G_1( M_{\tilde \ell} , \tilde \ell) = q \tilde \ell, \, \,  G_1( 0 , \tilde \ell) = 0. \label{eq:G1_limits}
\end{equation} 
So, $I_1$ is increasing with respect to $C$ and satisfies $I_1(0,\tilde \ell) = 0$ and 
\begin{equation} 
I_1(M_{\tilde \ell}, \tilde \ell)  = \int_{-1/2}^{1/2} G_1 (M_{\tilde \ell} e^{- \kappa' V(x_1)}, \tilde \ell) \, d x_1 = \tilde \ell {\mathcal I}_1, 
\label{eq:prop_C1_ltil}
\end{equation}
where ${\mathcal I}_1$ is given by \eqref{eq:I1} and we have used the property that $ G_1 (\tilde \ell y, \tilde \ell) = \tilde \ell G_1(y,1)$, which follows from the same property of $\tilde F$: $ \tilde F(\tilde \ell \rho, \tilde \ell) = \tilde \ell \tilde F(\rho, 1)$. So, there exists $C \in [0, M_{\tilde \ell}]$ such 
that $I_1(C,\tilde \ell) = 1$ if and only if $\tilde \ell \geq {\mathcal I}_1^{-1}$ and, if this condition is satisfied, there exists a unique such $C$ still denoted by $C_{\tilde \ell}$. We abbreviate $\rho^1_{C_{\tilde \ell},\tilde \ell}$ into $\rho_{\tilde \ell}$ again. Then, \eqref{eq:str_bet_2} leads to \eqref{eq:express_rho_1} for $u_2$ and with the fact that $\rho_{\tilde \ell} \leq q \tilde \ell < \tilde \ell$, we deduce that $u_2$ has the same sign as $\lambda$. With \eqref{eq:rho_ell_negative} (with $k=1$) and \eqref{eq:G1_limits}, we also get that $\rho_{\tilde \ell}$ is even, decreasing on $[0,1/2]$ and such that $\lim_{x_1 \to 1/2} \rho (x_1) = 0$. Since $|u_2| = \frac{|bZ|}{c_1} (\tilde \ell - \rho_{\tilde \ell})$, we get that $|u_2|$ is even, increasing on $[0,1/2]$ and such that $\lim_{x_1 \to 1/2} |u_2 (x_1)| = \frac{|bZ|}{c_1} \tilde \ell = \max_{[-1/2,1/2]} |u_2|$. The admissibility condition $\max_{[-1/2,1/2]} |u_2| \leq 1$ is thus satisfied if and only if $\tilde \ell \leq \frac{c_1}{|bZ|}$. Collecting the conditions on $\tilde \ell$, we get
$$ {\mathcal I}_1^{-1} \leq \tilde \ell \leq \frac{c_1}{|bZ|}. $$
Thus, the existence of $\tilde \ell$ requires Condition \eqref{eq:strip_exist_classb} to be satisfied. 

Then, we turn towards $u_1$. If $\tilde \ell < \frac{c_1}{|bZ|}$, then $|u_2|$ never reaches the value $1$ and so, $u_1$ cannot change sign and we have two opposite solutions for $u_1$, either positive or negative. If $\tilde \ell = \frac{c_1}{|bZ|}$, $u_2$ reaches the value $1$ but at the boundary of the interval. So, $u_1$ cannot change sign in the interior of the domain, and the situation is similar to the previous case: there are two opposite solution for $u_1$, either positive or negative in $(-1/2,1/2)$. The only difference is that $u_1$ vanishes at the boundary. 
Again, the properties of $\beta$ follow from integrating \eqref{eq:integration_beta'}. This completes the determination of Class (b) of solutions.

\medskip
\textbf{Class (c):} The definition of $\rho^2_{C,\tilde \ell}$ also requires $C \leq M_{\tilde \ell}$. We let $I_2(C, \tilde \ell)$ be defined by~\eqref{eq:def_I3} with $G_3$ substituted by $G_2$. The function $G_2$ satisfies: 
\begin{equation}
y \mapsto G_2( y , \tilde \ell) \textrm{ is decreasing on } [0,M_{\tilde \ell}], \, \,  G_2( M_{\tilde \ell} , \tilde \ell) = q \tilde \ell, \, \,  G_2( 0 , \tilde \ell) = \tilde \ell. \label{eq:G2_limits}
\end{equation} 
Therefore, $I_2$ is decreasing with respect to $C$ and satisfies
\begin{equation} 
I_2(0,\tilde \ell) = \tilde \ell \quad \textrm{and} \quad I_2(M_{\tilde \ell}, \tilde \ell)  = \int_{-1/2}^{1/2} G_2 (M_{\tilde \ell} e^{- \kappa' V(x_1)}, \tilde \ell) \, d x_1 = \tilde \ell {\mathcal I}_2, 
\label{eq:prop_C2_ltil}
\end{equation}
for the same reason as for $I_1$. We note that ${\mathcal I}_2 <1$ because $G_2(y,1) <1$ except for $y = 0$. Thus, existence of $C$ such that $I_2(C,\tilde \ell) = 1$ requires 
\begin{equation}
1 \leq \tilde \ell \leq {\mathcal I}_2^{-1}. 
\label{eq:rho2_cond_on_ltil}
\end{equation}
Under this condition, there exists a unique such $C$ still denoted by $C_{\tilde \ell}$ and we also abbreviate $\rho^2_{C_{\tilde \ell},\tilde \ell}$ into $\rho_{\tilde \ell}$. With \eqref{eq:G2_limits}, $\rho_{\tilde \ell}$ is even, increasing on $[0,1/2)$ and such that $\lim_{x_1 \to 1/2} \rho_{\tilde \ell} = \tilde \ell$. Then, $u_2$ is given by \eqref{eq:express_rho_2} and $u_2$ has the sign of $\lambda$. Furthermore $|u_2|$ is even, decreasing on $[0,1/2)$ and such that $\lim_{x_1 \to 1/2} |u_2| = 0$. So, $|u_2|$ is maximal at $x_1 = 0$. Then, $|u_2|$ is admissible provided that 
\begin{equation}
\max_{(-1/2,1/2)} |u_2| = |u_2(0)| = \frac{|bZ|}{c_1} H_2(\tilde \ell)  \leq 1, \, \, \textrm{ with } \, \, H_2(\tilde \ell) = \tilde \ell - G_2(C_{\tilde \ell}, \tilde \ell), 
\label{eq:admiss_class_c}
\end{equation}
where $H_2(\tilde \ell)$ is defined for $\tilde \ell \in [1, {\mathcal I}_2^{-1}]$. 

Thanks to \eqref{eq:prop_C2_ltil}, we have $I_2(0,1) = 1$ and $I_2(M_{{\mathcal I}_2^{-1}}, {\mathcal I}_2^{-1}) = 1$. This implies that $C_1 = 0$ and $C_{{\mathcal I}_2^{-1}} = M_{{\mathcal I}_2^{-1}}$, so that $G_2(C_1,1) = 1$ and $G_2(C_{{\mathcal I}_2^{-1}}, {\mathcal I}_2^{-1}) = q {\mathcal I}_2^{-1}$. It results that 
$$ H_2(1) = 0, \qquad H_2 ({\mathcal I}_2^{-1}) = (1-q) {\mathcal I}_2^{-1}.$$ 
We now show that $H_2$ is increasing on $[1, {\mathcal I}_2^{-1}]$. The proof is slightly more involved than in the cases of Class (a) or $\ell >0$ because we will need to compute $\frac{d C_{\tilde \ell}}{d \tilde \ell}$. Similar computations as before show that 
\begin{equation}
H_2' ( \tilde \ell) = \frac{\tilde \rho^{1-q} (\tilde \ell - \tilde \rho)^q}{\tilde \rho - q \tilde \ell} \Big[ \frac{d C_{\tilde \ell}}{d \tilde \ell} - q \Big( \frac{\tilde \ell}{\tilde \rho} - 1 \Big)^{1-q} \Big], \, \, \textrm{ with } \, \, \tilde \rho = G_2(C_{\tilde \ell},\tilde \ell). 
\label{eq:H2prime}
\end{equation}
We compute 
$$ \frac{d C_{\tilde \ell}}{d \tilde \ell} (\tilde \ell) = (1-q) \frac{\displaystyle \int_{-1/2}^{1/2} \Big( \frac{\rho(x_1)}{\tilde \ell - \rho(x_1)} \Big)^q \, \frac{dx_1}{D\big(\rho(x_1)\big)}}{\displaystyle \int_{-1/2}^{1/2} e^{-\kappa' V(x_1)} \,  \frac{dx_1}{D\big(\rho(x_1)\big)}}, $$
with $ \rho(x_1) = G_2(C_{\tilde \ell} e^{-\kappa' V(x_1)},\tilde \ell)$ and 
$$ D(\rho) = (1-q) \Big( \frac{\rho}{\tilde \ell - \rho} \Big)^q - q \Big( \frac{\tilde \ell - \rho}{\rho} \Big)^{1-q} = - \frac{\partial F}{\partial \rho} (\rho, \tilde \ell) \geq 0. $$

Suppose that $0 < C_{\tilde \ell} < M_{\tilde \ell}$, i.e. $1 < \tilde \ell < {\mathcal I}_2^{-1}$, so that $ q \tilde \ell < \tilde \rho < \tilde \ell$. We note that $D$: $[q \tilde \ell, \tilde \ell) \to [0,\infty)$ is increasing and onto. Thus, as $\rho$ is minimal at $x_1 = 0$ with value $\rho(0) = \tilde \rho$, we get 
$$ D\big(\rho(x_1)\big) \geq D(\tilde \rho) > 0, \quad \forall x_1 \in (-1/2,1/2).  $$
Then, we have 
$$\int_{-1/2}^{1/2} e^{-\kappa' V(x_1)} \,  \frac{dx_1}{D\big(\rho(x_1)\big)} \leq \int_{-1/2}^{1/2}  \,  \frac{dx_1}{D\big(\rho(x_1)\big)} \leq \frac{1}{D(\tilde \rho)} < \infty. $$
Furthermore, since the function $[q \tilde \ell, \tilde \ell) \to [\frac{q}{1-q},\infty)$, $\rho \to \frac{\rho}{\tilde \ell - \rho}$ is increasing, we have 
$$ \frac{\rho(x_1)}{\tilde \ell - \rho(x_1)} \geq \frac{\tilde \rho}{\tilde \ell - \tilde \rho} . $$
So, 
$$ \int_{-1/2}^{1/2} \Big( \frac{\rho(x_1)}{\tilde \ell - \rho(x_1)} \Big)^q \, \frac{dx_1}{D\big(\rho(x_1)\big)} \geq \Big( \frac{\tilde \rho}{\tilde \ell - \tilde \rho} \Big)^q \int_{-1/2}^{1/2}  \,  \frac{dx_1}{D\big(\rho(x_1)\big)}. $$
It follows that 
$$ \frac{d C_{\tilde \ell}}{d \tilde \ell} (\tilde \ell) \geq (1-q) \Big( \frac{\tilde \rho}{\tilde \ell - \tilde \rho} \Big)^q, $$
so that 
\begin{equation}
\frac{d C_{\tilde \ell}}{d \tilde \ell} (\tilde \ell) -  q \Big( \frac{\tilde \ell}{\tilde \rho} - 1 \Big)^{1-q} \geq D(\tilde \rho) > 0, \quad \forall \tilde \ell \in (1,{\mathcal I}_2^{-1}). 
\label{eq:Ctillprime}
\end{equation}
From this and \eqref{eq:H2prime}, it follows that $H_2'(\tilde \ell) > 0$, for all $\tilde \ell \in (1, {\mathcal I}_2^{-1})$, which shows that $H_2$ is increasing on $[1, {\mathcal I}_2^{-1}]$.

We see that if condition \eqref{eq:cond_class_c} is satisfied, any $\tilde \ell \in [1,{\mathcal I}_2^{-1}]$ gives rise to a solution satisfying the admissibility condition \eqref{eq:admiss_class_c}. On the other hand, if \eqref{eq:cond_class_c} is not satisfied, we must restrict the range of admissible $\tilde \ell$ to $[1, \tilde \ell^*_c]$ with $\tilde \ell^*_c$ being the unique solution of $\frac{|bZ|}{c_1} H_2(\tilde \ell^*) = 1$. Once $u_2$ is found, the derivations of $u_1$ and $\beta$ follow the same steps as in Class (a) or as in the case $\ell >0$. 

\medskip
\textbf{Class (d):} The conditions under which $\rho^4_{\tilde \ell}$ and $\rho^5_{\tilde \ell}$ give rise to a travelling-wave solution are the same. So, w.l.o.g., we restrict ourselves to $\rho^4$. Using the same ideas as for Class~(b), the integral of $\rho^4_{\tilde \ell}$ is given by 
\begin{eqnarray*}
 I_4(\tilde \ell) &=& \int_{-1/2}^0 G_1(M_{\tilde \ell} e^{- \kappa' V(x_1)}, \tilde \ell) \, dx_1 + \int_0^{1/2} G_2(M_{\tilde \ell} e^{- \kappa' V(x_1)}, \tilde \ell) \, dx_1 \\
&=& \tilde \ell \, {\mathcal I}_{12}. 
\end{eqnarray*}
Thus, the normalization condition \eqref{eq:rho_normaliz_strip} is satisfied if and only if $\tilde \ell = {\mathcal I}_{12}^{-1}$ and the corresponding solution is denoted by $\rho_{\tilde \ell}$ for simplicity. Thus, $\rho_{\tilde \ell}$ is increasing on $[-1/2, 1/2]$ with $\lim_{x_1 \to -1/2} \rho(x_1) = 0$, $\rho(0) = q \tilde \ell$, $\lim_{x_1 \to 1/2} \rho(x_1) = \ell$. The function $u_2$ is given by \eqref{eq:express_rho_4} and has the sign of $\lambda$. Furthemore, $|u_2| = \frac{|bZ|}{c_1} (\tilde \ell - \rho_{\tilde \ell})$ is decreasing on $[-1/2, 1/2]$ with $\lim_{x_1 \to -1/2} |u_2|(x_1) = \frac{|bZ|}{c_1} \tilde \ell$, $|u_2|(0) = \frac{|bZ|}{c_1} (1-q) \tilde \ell$, $\lim_{x_1 \to 1/2} |u_2|(x_1) = 0$. Thus, $u_2$ is an admissible solution if and only if $\frac{|bZ|}{c_1} \tilde \ell \leq 1$, or equivalently \eqref{eq:cond_class_d}. If this condition is satisfied, it is clear that there are two solutions for $u_1$, one positive, one negative. Similarly, the corresponding solutions for $\beta$ are opposite one to another.  This ends the discussion of this case and the proof of Prop.\ref{prop:strip_negative}.  \endproof

\setcounter{equation}{0}
\section{Proofs (annular geometry case)}
\label{sec_rotat_sym_proof}

In this section, we give a proof of Prop.~\ref{prop:rotat_sym}. Inserting \eqref{eq:sol_annul_form} in the NSH system \eqref{eq:fl_rho_sm}-\eqref{eq:fl_al_sm}, we get
\begin{eqnarray}
&&\hspace{-1cm}
\big[ r \rho \, \big( c_1 u_r + b \rho \, \beta' \big) \big]' = 0, \label{eq:fl_rho_rad} \\
&&\hspace{-1cm}
\big( c_2 u_r  + b \rho \,\beta' \big) \, u_r' - \frac{u_\vartheta}{r} \big( c_2 u_\vartheta +  \frac{bm}{r} \rho \big) + u_\vartheta^2 \, \big[ \Theta \log \rho + \kappa V \big]' = 0,  \label{eq:fl_ur_rad} \\
&&\hspace{-1cm}
\big( c_2 u_r + b \rho \, \beta' \big) \, u_\vartheta' + \frac{u_r}{r} \big( c_2 u_\vartheta + \frac{bm}{r} \rho \big) -  u_r u_\vartheta \, \big[ \Theta \log \rho + \kappa V \big]'  = 0,  \label{eq:fl_uth_rad} \\
&&\hspace{-1cm}
- m \lambda + \big( c_1 u_r + b \rho \, \beta' \big) \, \beta' + \frac{m}{r} \big( c_1 u_\vartheta +  \frac{bm}{r} \rho \big) = 0, \label{eq:fl_al_rad}
\end{eqnarray}
where primes denote derivatives with respect to $r$. Compared with the cartesian case, we note the presence of extra terms stemming from inertia forces. Like in the cartesian case, we remark that \eqref{eq:fl_uth_rad} is a consequence of \eqref{eq:fl_ur_rad} and of \eqref{eq:str_const_cyl} (wherever $u_\vartheta \not = 0$) and can be ignored. We also notice that \eqref{eq:fl_rho_rad} is implied by the condition 
\begin{equation}
c_1 u_r + b \rho \, \beta' = 0.  
\label{eq:zero_current}
\end{equation}
which means that the radial velocity vanishes. From now on, we restrict ourselves to solutions which satisfy \eqref{eq:zero_current}. 

Since $m \not = 0$, from \eqref{eq:fl_al_rad} we get 
\begin{equation}
c_1 u_\vartheta + \frac{bm}{r} \rho = \lambda r.
\label{eq:utheta}
\end{equation} 
We note that \eqref{eq:str_const_cyl} implies 
\begin{equation} 
u_r \, u_r' = - u_\vartheta \, u_\vartheta', 
\label{eq:udu_cyl}
\end{equation}
Inserting \eqref{eq:zero_current} and \eqref{eq:utheta} into \eqref{eq:fl_ur_rad} and using \eqref{eq:udu_cyl}, we get 
\begin{equation} 
(c_1-c_2) \Big( u_\vartheta' + \frac{u_\vartheta}{r} \Big) + u_\vartheta \big[ \Theta \log \rho + \kappa V \big]' = \lambda, 
\label{eq:utheta_prime}
\end{equation}
wherever $u_\vartheta \not = 0$. With $\ell$, $q$ and $\kappa'$ respectively given by \eqref{eq:annular_def_ell} and \eqref{eq:annul_calI} and recalling that $\ell \geq 0$, Eqs. \eqref{eq:utheta} and \eqref{eq:utheta_prime} can be rewritten
\begin{eqnarray}
&& \hspace{-1cm}
u_\vartheta = - \frac{bm}{c_1 r} (\rho + \ell r^2), \label{eq:utheta1}\\
&& \hspace{-1cm}
u_\vartheta \Big[ (1-q) \Big( \frac{u_\vartheta'}{u_\vartheta} + \frac{1}{r} \Big) + q \frac{\rho'}{\rho}  + \kappa' V' \Big] = \frac{\lambda}{c_1-c_2 + \Theta}. \label{eq:utheta_prime1}
\end{eqnarray}
Note that \eqref{eq:utheta1} shows that $u_\vartheta$ indeed never vanishes. 

Inserting \eqref{eq:utheta1} into \eqref{eq:utheta_prime1} to eliminate $u_\vartheta$, we get 
\begin{equation}
\rho' = \frac{\rho}{\rho + q \ell r^2} F_\ell(r,\rho), \qquad 
\label{eq:ann_rho}
\end{equation}
with 
$$ F_\ell(r,\rho) = a \ell r - (\rho + \ell r^2) \kappa' V'(r) \qquad \textrm{and} \qquad a = \frac{2c_2 - c_1}{c_1-c_2 + \Theta}, $$
with $a>0$ thanks to \eqref{eq:2c2-c1}. The function $F_\ell$ is defined for $(r,\rho) \in {\mathcal O} =: (R_0,R_1) \times [0,\infty)$. It can be equivalently written
\begin{equation} 
F_\ell(r,\rho) = - \kappa' V'(r) \big( \rho - \tilde \rho_\ell(r) \big), \qquad \tilde \rho_\ell(r) = \ell r \Big( \frac{a}{\kappa' V'(r)} - r \Big). 
\label{eq:def_til_rho}
\end{equation}
From the hypotheses of Prop.~\ref{prop:rotat_sym}, we recall that $R^* \in (R_0,R_1)$ is uniquely defined by  $V(R^*) = \min_{r \in (R_0,R_1)} V(r)$ and that $V'<0$ on $(R_0,R^*)$ and $V'>0$ on $(R^*,R_1)$. 

\begin{figure}[ht!]
\centering

$\mbox{}$ \hspace{0.cm}
\subfloat[]{\includegraphics[trim={4cm 15cm 5.5cm 4.5cm},clip,height= 5cm]{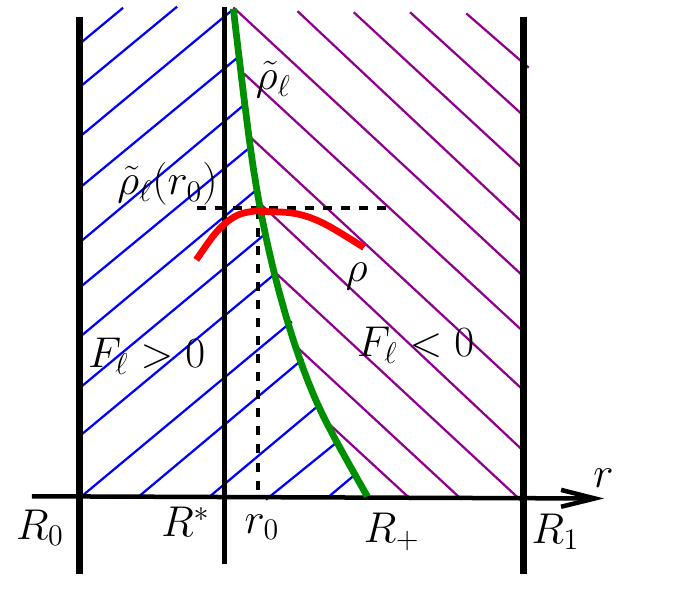}\label{subfig:annulus_rho_construct}} \hspace{0.5cm}
\subfloat[]{\includegraphics[trim={4cm 15cm 5.5cm 4.5cm},clip,height= 5cm]{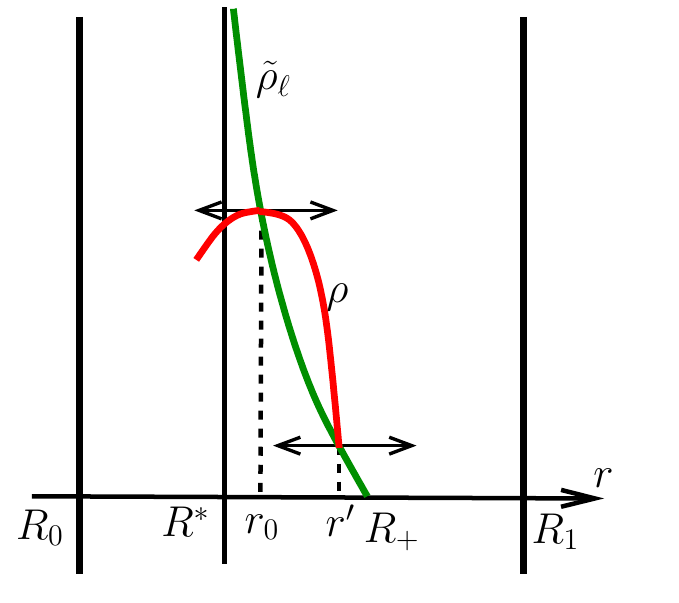}\label{subfig:annulus_rho_obstruction}} 

\caption{(a): construction of the solution $\rho_{r_0, \ell}$ (in red) having Cauchy data $(r, \rho_{r_0, \ell}^*)$ with $\rho_{r_0, \ell}^* = \tilde \rho_\ell(r_0) >0$. The function $r \mapsto \tilde \rho_\ell(r)$ is drawn in green. Due to the sign of $F_\ell$, $(r, \rho_{r_0, \ell}^*)$ is a local maximum of $\rho_{r_0, \ell}$. (b): obstruction to a change of monotony of $\rho_{r_0, \ell}$ at a point $r'>r_0$. If $r'$ is the first of these monotony changes, $\rho_{r_0, \ell}'(r')$ should be zero, but it cannot be as $\rho_{r_0, \ell}$ must be stricly decreasing between $r_0$ and $r'$.}
\label{fig:annulus_rho_construct}
\end{figure}

Prop.~\ref{prop:rotat_sym} is about the case $\ell >0$ so we solve \eqref{eq:ann_rho} in this case. We first note that there exists a unique $R_+ \in (R^*,R_1)$ such that  $\frac{a}{\kappa' V'(r)} -r = 0$. Indeed, this is equivalent to $r V'(r) = \frac{a}{\kappa'}$. But the function $[R^*,R_1) \to [0,\infty)$, $r \mapsto r V'(r)$ is strictly increasing, takes the value $0$ at $R^*$ and tends to $+\infty$ as $r \to R_1$, which shows the existence of a unique $R_+>R^*$ satisfying this equation. Then, we have $\tilde \rho_\ell(r) \geq 0$ if and only if $R^*<r \leq R_+$.  Thanks to the assumption that $\frac{V'}{r}$ is strictly increasing on $[R^*,R_1)$, the function $\tilde \rho_\ell$ is strictly decreasing on $(R^*,R_+]$, with $\tilde \rho_\ell \to \infty$ when $r \to R^*$ and $\tilde \rho_\ell(R_+) = 0$. Let $\Sigma$ be the representative curve of $\tilde \rho_\ell$ in ${\mathcal O}$ i.e. 
$$ \Sigma = \{ (r,\rho) \in (R^*,R_+) \times [0,\infty) \, \, | \, \, \rho = \tilde \rho_\ell(r) \}, $$
(see Fig. \ref{subfig:annulus_rho_construct} for a graphical representation of $\Sigma$). 
Then, ${\mathcal O} \setminus \Sigma$ is partitioned in two disjoint domains ${\mathcal O}_+$ and ${\mathcal O}_-$ with 
\begin{eqnarray*}
{\mathcal O}_+ &=&  \big( (R_0,R^*) \times [0,\infty) \big) \cup \{ (r,\rho) \in (R^*,R_+) \times [0,\infty) \, \, | \, \, \rho < \tilde \rho_\ell(r) \}, \\
{\mathcal O}_- &=& \big( (R_+,R_1) \times [0,\infty) \big) \cup\{ (r,\rho) \in (R^*,R_+) \times [0,\infty) \, \, | \, \, \rho > \tilde \rho_\ell(r) \}, 
\end{eqnarray*}
and the following is readily checked: 
\begin{equation}
\left\{ \begin{array}{ccc} F_\ell(r,\rho) >0 \, & \, \Longleftrightarrow \, & \, (r,\rho) \in {\mathcal O}_+, \\
F_\ell(r,\rho) =0 \, & \, \Longleftrightarrow \, & \, (r,\rho) \in \Sigma, \\
F_\ell(r,\rho) <0 \, & \, \Longleftrightarrow \, & \, (r,\rho) \in {\mathcal O}_- . \end{array} \right. 
\label{eq:signF}
\end{equation}

Now, we are going to solve \eqref{eq:ann_rho} with an initial condition on $\Sigma$ in the case $\ell >0$. Let $r_0 \in (R^*,R_+)$ and denote by $\rho_{r_0, \ell}^* = \tilde \rho_\ell(r_0) >0$. The right-hand side of \eqref{eq:ann_rho} is continuous and uniformly Lipschitz with respect to $\rho$ on all compact subsets of ${\mathcal O}$ of the form $[R_0+\epsilon, R_1 - \epsilon] \times [0,M]$ for all $\epsilon$ in a neighborhood $(0,\epsilon_0)$ of $0$ and all $M>0$. By the Cauchy-Lipschitz theorem, it results that, for any $r_0 \in (R^*,R_+)$, Eq. \eqref{eq:ann_rho} with Cauchy datum $(r_0, \rho_{r_0, \ell}^*)$ can be uniquely solved in an interval $(r_1,r_2)$ containing $r_0$. This defines a solution named $\rho_{r_0, \ell}$ (see a graphical representation of this construction in Fig. \ref{subfig:annulus_rho_construct}).

With the sign conditions \eqref{eq:signF}, $\rho_{r_0, \ell}$ is strictly increasing on $(r_1,r_0)$ and strictly decreasing on $(r_0,r_2)$. Indeed, this is certainly true in a small neighborhood of $r_0$. Now, suppose changes of monotony occur and denote by $r' \not = r_0$ the closest point to $r_0$ where this occurs. Then, $\rho_{r_0, \ell}'(r')=0$. Hence $\rho_{r_0, \ell}(r')= \tilde \rho_\ell(r')$. Since $\tilde \rho_\ell$ is decreasing, this is only possible if $r'>r_0$. Indeed, if $r' < r_0$, we have $\tilde \rho_\ell(r') > \rho_{r_0, \ell}^*$ while $\rho_{r_0, \ell}(r') < \rho_{r_0, \ell}^*$, which is a contradiction since changes of monotony can only occur on the graph $\Sigma$ of $\tilde \rho_\ell$. Now, $(r',\rho_{r_0, \ell}(r')) \in \Sigma$ and the sign conditions \eqref{eq:signF} impose $\rho_{r_0, \ell}'(r) > 0$ for $r$ in a left neighborhood of $r'$. On the other hand, $\rho_{r_0, \ell}'(r)<0$ in $(r_0,r')$ because it is so in a right neighborhood of $r_0$ and as long as $r$ does not meet the first monotony change~$r'$. This leads to a contradiction (see a graphical representation of this obstruction in Fig. \ref{subfig:annulus_rho_obstruction}) and shows that $\rho_{r_0, \ell}$ is strictly increasing on $(r_1,r_0)$ and strictly decreasing on $(r_0,r_2)$. Note that $r_1$ and $r_2$ may depend on $r_0$ and $\ell$. 

What precedes proves that $\rho_{r_0, \ell} \leq \rho_{r_0, \ell}^*$. Thus, $\rho_{r_0, \ell}$ can be extended to the left of $r_1$ or to the right of $r_2$ as long as it does not take the value $0$. Suppose there exists $\tilde r \in (R_0,R_1)$ such that $\rho_{r_0, \ell}(\tilde r) = 0$. Then, inspection of  \eqref{eq:ann_rho} shows that there is a unique solution with Cauchy datum $(\tilde r, 0)$ and this solution is identically zero. By the uniqueness part of the Cauchy-Lipshitz theorem, $\rho_{r_0, \ell}$ must coincide with this solution and hence must be identically zero, which is in contradiction with the fact that $\rho_{r_0, \ell}^*>0$. Hence $\rho_{r_0, \ell}$ is defined on the whole interval $(R_0,R_1)$. 

Now, we show that, for all $\ell >0$, $r_0 \in (R^*,R_+)$ and $r \in (R_0,R_1)$, we have 
\begin{equation}
C^{-2} \, \rho_{r_0,\ell}^* \, e^{-\frac{\kappa' V(r)}{q}} \leq \rho_{r_0,\ell}(r) \leq C \, \rho_{r_0,\ell}^* \, e^{-\kappa' V(r)}, 
\label{eq:sol_estim}
\end{equation}
with $C = e^{\frac{a(R_1-R_0)}{qR_0}}$. Indeed, from \eqref{eq:ann_rho}, we can write (we abbreviate $\rho_{r_0,\ell}$ into $\rho$ for simplicity): 
$$ \frac{\rho'}{\rho} = \frac{a \ell r}{\rho + q \ell r^2} - \frac{\rho + \ell r^2}{\rho + q \ell r^2} \kappa' V'(r), $$
and we have on ${\mathcal O}$: 
$$ 0 < \frac{a \ell r}{\rho + q \ell r^2}  \leq \frac{a}{q R_0}, \qquad 1 < \frac{\rho + \ell r^2}{\rho + q \ell r^2} \leq \frac{1}{q}. $$
Now, suppose $r > R^*$. Then, 
$$ - \frac{\kappa' V'(r)}{q}  \leq \frac{\rho'}{\rho} \leq  \frac{a}{q R_0} - \kappa' V'(r).$$  
Integrating this inequality on $[R^*,r]$ and remembering that $V(R^*) = 0$, we get:
\begin{equation} 
\rho(R^*) e^{-\frac{\kappa' V(r)}{q}} \leq \rho (r) \leq C \rho(R^*) e^{-\kappa' V(r)}, \qquad \forall r \in [R^*,R_1).  
\label{eq:ineq1}
\end{equation}
Now, we know that $\rho(R^*) \leq \rho_{r_0,\ell}^*$. Furthermore, applying \eqref{eq:ineq1} at $r=r_0$, we also have 
$$  C^{-1} \rho_{r_0,\ell}^* \leq  C^{-1} \rho_{r_0,\ell}^* e^{\kappa' V(r_0)} \leq \rho(R^*). $$
Thus, we can eliminate $\rho(R^*)$ from \eqref{eq:ineq1} in favor of $\rho_{r_0,\ell}^*$ and get 
\begin{equation} 
C^{-1} \rho_{r_0,\ell}^* e^{-\frac{\kappa' V(r)}{q}} \leq \rho (r) \leq C \rho_{r_0,\ell}^* e^{-\kappa' V(r)}, \qquad \forall r \in [R^*,R_1).  
\label{eq:ineq1-2}
\end{equation}
For $r<R^*$, we similarly have 
$$ - \kappa' V'(r) \leq \frac{\rho'}{\rho} \leq \frac{a}{q R_0} - \frac{\kappa' V'(r)}{q}.   $$
Integrating this inequality on $[r,R^*]$, we get 
$$ \rho(R^*) C^{-1} e^{-\frac{\kappa' V(r)}{q}} \leq \rho (r) \leq \rho(R^*) e^{-\kappa' V(r)}, \qquad \forall r \in (R_0,R^*]. $$ 
and again, eliminating $\rho(R^*)$  in favor of $\rho_{r_0,\ell}^*$, we get 
\begin{equation} 
C^{-2} \rho_{r_0,\ell}^* e^{-\frac{\kappa' V(r)}{q}} \leq \rho (r) \leq \rho_{r_0,\ell}^* e^{-\kappa' V(r)}, \qquad \forall r \in (R_0,R^*].  
\label{eq:ineq2}
\end{equation}
Since $C>1$, we see that \eqref{eq:ineq1-2} and \eqref{eq:ineq2} imply \eqref{eq:sol_estim}. As a consequence of \eqref{eq:sol_estim}, we have 
\begin{equation}
\rho_{r_0,\ell}(r) \to 0 \quad \textrm{ as } \quad  r \to R_0 \quad \textrm{or} \quad r \to R_1. 
\label{eq:limrho}
\end{equation}
Indeed, this follows from the fact that $V(r) \to + \infty$ in these limits. 

In summary, for all $r_0 \in (R^*,R_+)$ there exists a unique solution $\rho_{r_0, \ell}$ to \eqref{eq:ann_rho} such that $\rho_{r_0, \ell}(r_0) = \rho_{r_0, \ell}^* = \tilde \rho_\ell(r_0)$. This solution is defined on $(R_0, R_1)$, tends to $0$ as $r \to R_0$ or $R_1$, is strictly increasing on $(R_0, r_0)$, strictly decreasing on $(r_0,R_1)$, and 
\begin{equation}
\max_{(R_0,R_1)} \rho_{r_0, \ell} = \rho_{r_0, \ell}(r_0) = \rho_{r_0, \ell}^* = \tilde \rho_\ell(r_0). 
\label{eq:max_rho}
\end{equation}
Such a solution is depicted in Fig. \ref{fig:annulus_rhoV}. 

We now show that there exists a unique value of $r_0 \in (R^*,R_+)$ such that the associated solution $\rho_{r_0, \ell}$ satisfies the normalization condition \eqref{eq:rho_normaliz}. Define
$$ I(r_0,\ell) = 2 \pi \int_{R_0}^{R_1} \rho_{r_0, \ell} (r) \, r \, dr. $$
We want to show that there exists a unique $r_0$ such that $I(r_0,\ell)=1$. First, using \eqref{eq:sol_estim}, we have for all $r_0 \in (R^*, R_+)$ and all $\ell >0$:  
\begin{equation} 
C^{-2} \rho_{r_0,\ell}^* {\mathcal I}_q \leq I(r_0,\ell) \leq C \rho_{r_0,\ell}^* {\mathcal I}_1,
\label{eq:estim_I}
\end{equation}
with 
$${\mathcal I}_1 = 2 \pi \int_{R_0}^{R_1} e^{-\kappa' V(r)} \, r \, dr, \qquad {\mathcal I}_q = 2 \pi \int_{R_0}^{R_1} e^{-\frac{\kappa' V(r)}{q}} \, r \, dr. $$
The integrals ${\mathcal I}_q$ and ${\mathcal I}_1$ are just constants and we note that ${\mathcal I}_q \leq {\mathcal I}_1$. Now, since $\rho_{r_0,\ell}^* \to 0$ when $r_0 \to R_+$ and $\rho_{r_0,\ell}^* \to + \infty$ when $r_0 \to R^*$, we have $I(r_0,\ell) \to 0$ when $r_0 \to R_+$ and $I(r_0,\ell) \to + \infty$ when $r_0 \to R^*$. 

Then, we note that 
\begin{equation}
\big( r_1 < r_2 \big) \quad \Longrightarrow \quad \big( \rho_{r_1,\ell}(r) > \rho_{r_2,\ell}(r), \quad \forall r \in (R_0,R_1) \big).
\label{eq:compar}
\end{equation}
Indeed, if $r_1 < r_2$, we have 
$$ \rho_{r_1,\ell}(r_1) = \rho_{r_1,\ell}^* > \rho_{r_2,\ell}^* = \rho_{r_2,\ell}(r_2) > \rho_{r_2,\ell}(r_1). $$
Thus, if \eqref{eq:compar} was untrue, by the continuity of $\rho_{r_1,\ell}$, $\rho_{r_2,\ell}$, there would exist $r$ such that $\rho_{r_1,\ell}(r) = \rho_{r_2,\ell}(r)$. But then, by the uniqueness in the Cauchy-Lipschitz theorem, these two solutions would be equal, which would imply $r_1=r_2$, leading to a contradiction. From \eqref{eq:compar}, we deduce that $I$ is strictly decreasing with respect to $r_0$. 

We also have that $I$ is continuous with respect to $r_0$. Indeed, it is an easy matter to see that the right-hand side of \eqref{eq:ann_rho} is a function $G(r,\rho,\ell)$ which belongs to $C^1\big((R_0,R_1) \times (0,\infty) \times (0,\infty)\big)$. Now, the function $(r,\ell) \mapsto \tilde \rho_\ell(r)$ defined by \eqref{eq:def_til_rho} belongs to $C^1\big( (R^*,R_+) \times (0,\infty) \big)$ as well. Hence, by the smooth dependency of the solution of a differential equation upon initial conditions and parameters \cite{dieudonne2011foundations}, we get that the function $(r,r_0,\ell) \mapsto \rho_{r_0,\ell}(r)$ belongs to $C^0\big((R_0,R_1) \times (0,\infty) \times (0,\infty)\big)$. Now, by \eqref{eq:max_rho} and the continuity of $\tilde \rho_\ell$, for any compact intervals $K \subset (R^*,R_+)$ and $L \subset (0,\infty)$, there exists a constant $M>0$ such that $\rho_{r_0,\ell}(r) \leq M$, for all $(r,r_0,\ell) \in (R_0,R_1) \times K \times L$. By the dominated convergence theorem, it follows that $I$ is continuous on $(R^*,R_+) \times (0,\infty)$. Hence, for every $\ell \in (0,\infty)$, there exists a unique $r_\ell \in (R^*, R_+)$ such that 
\begin{equation}
I(r_\ell, \ell) = 1. 
\label{eq:def_rl}
\end{equation}
Furthermore, the function $\ell \to r_\ell$ is in $C^0\big((0,\infty)\big)$. We denote the corresponding solution by $\rho_\ell = \rho_{r_\ell,\ell}$ and its maximum value by $\rho_\ell^* = \rho_{r_\ell,\ell}^* = \tilde \rho_\ell(r_\ell)$. 

We note that the function $(r,\ell) \mapsto \rho_\ell(r)$ belongs to $C^0 \big( (R_0,R_1) \times (0,\infty) \big)$. 
Since $\rho_\ell(r) \to 0$ as $r \to R_0$ and $r \to R_1$, we extend $\rho_\ell$ by continuity into a function defined  on $[R_0,R_1]$ which takes the value $0$ at $R_0$ and $R_1$. We now show that the so-extended function $(r,\ell) \mapsto \rho_\ell(r)$  belongs to $C^0 \big( [R_0,R_1] \times (0,\infty) \big)$. This amounts to show that $\rho_{\ell}(r) \to \rho_{\ell_0}(R_0) = 0$ as $(r,\ell) \to (R_0,\ell_0)$, for any $\ell_0 >0$, and similarly with $R_1$. First, with \eqref{eq:estim_I} and \eqref{eq:def_rl}, we have 
\begin{equation} 
\frac{1}{C {\mathcal I}_1} \leq \rho_\ell^* \leq \frac{C^2}{{\mathcal I}_q}. 
\label{eq:estim_rho*}
\end{equation}
Inserting this into \eqref{eq:sol_estim} with $r_0= r_\ell$, we get for all $r \in (R_0,R_1)$
\begin{equation}
\frac{1}{C^3 {\mathcal I}_1}  \, e^{-\frac{\kappa' V(r)}{q}} \leq \rho_\ell(r) \leq \frac{C^3}{{\mathcal I}_q} \, e^{-\kappa' V(r)}.
\label{eq:estim_rho}
\end{equation} 
From this inequality, we clearly get the requested convergence. 

We now return to \eqref{eq:utheta1} and look for sufficient conditions on $\ell$ such that the constraint 
\begin{equation}
|u_\vartheta| \leq 1,
\label{eq:constraint_utheta}
\end{equation}
is satisfied. We define 
\begin{equation}
v = - \frac{c_1}{bm} u_\vartheta = \frac{\rho}{r} + \ell r, 
\label{eq:defv}
\end{equation}
and denote by $v_\ell$ the function $v$ associated to $\rho_\ell$ through \eqref{eq:defv}. 
We note that $v_\ell \geq \ell r > 0$ so that $u_\vartheta$ is of constant sign, given by the sign of $-bm$. Like $\rho_\ell$, $v_\ell$ is defined and continuous on $[R_0,R_1]$ and the function $(r,\ell) \mapsto v_\ell(r)$ belongs to $C^0 \big( [R_0,R_1] \times (0,\infty) \big)$. We denote by $v_\ell^* = \max_{(R_0,R_1)} v_\ell <\infty$. The constraint \eqref{eq:constraint_utheta} translates into 
\begin{equation}
 v_\ell^* \leq \frac{c_1}{|bm|}. 
\label{eq:constraint_vell}
\end{equation}

Since the function $(r,\ell) \mapsto v_\ell(r)$ belongs $C^0 \big( [R_0,R_1] \times (0,\infty) \big)$, it is uniformly continuous on all compact sets of the form $[R_0,R_1] \times [\ell_1,\ell_2]$, with $0 < \ell_1 < \ell_2$. Consequently, the map $\ell \mapsto v_\ell^*$ belongs to $C^0\big((0,\infty)\big)$. We now show that this map is continuous at $\ell = 0$. This requires first to define $\rho_0$ and $v_0$, which has not been done so far. If $\ell = 0$, Eq. \eqref{eq:ann_rho} is easily solved together with the normalization condition \eqref{eq:rho_normaliz} and gives 
\begin{equation} 
\rho_0(r) = \frac{1}{{\mathcal I}_1} e^{- \kappa' V(r)}, \quad \forall r \in (R_0,R_1). 
\label{eq:def_rho0}
\end{equation}
We note that $\rho_0^* =: \max_{(R_0,R_1)} \rho_0 = \frac{1}{{\mathcal I}_1}$ and the maximum is attained at the point $r = R^*$. We also note that $\rho_0$ is the only solution to Eq. \eqref{eq:ann_rho} (with $\ell = 0$) satisfying \eqref{eq:rho_normaliz}. 

We now show that $\rho_\ell(r) \to \rho_0(r)$ as $\ell \to 0$ uniformly on $[R_0,R_1]$. First, \eqref{eq:estim_rho*} shows that $\rho_\ell^*$ stays in a compact subset of $(0,\infty)$ as $\ell$ ranges in $(0,\infty)$. Thus, for any sequence $\ell_n \to 0$, there exists a subsequence still denoted by $\ell_n$ for simplicity, and $\bar \rho^* >0$ such that $\rho_{\ell_n}^* = \tilde \rho_{\ell_n}(r_{\ell_n}) \to \bar \rho^*$ as $n \to \infty$. With \eqref{eq:def_til_rho} and abbreviating $r_{\ell_n}$ into $r_n$, $\rho_{\ell_n}$ into $\rho_n$, $\rho_{\ell_n}^*$ into $\rho_n^*$, this implies that 
\begin{equation} 
r_n \Big( \frac{a}{\kappa' V'(r_n)} - r_n \Big) =  \frac{\rho_n^*}{\ell_n} \to + \infty, \quad \textrm{ as } \quad n \to \infty. 
\label{eq:lim_rn}
\end{equation}
Now, the function at the left-hand side of \eqref{eq:lim_rn} (replacing $r_n$ by an arbitrary $r$) is continuous and strictly decreasing with respect to $r \in (R^*,R_+)$ and tends to $+ \infty$ when $r \to R^*$. It follows that $r_n \to R^*$. Thus, the Cauchy datum $(r_n,\rho_n^*)$ which defines the solution $\rho_n$ of \eqref{eq:ann_rho} for $\ell = \ell_n$ converges to the Cauchy datum $(R^*, \bar \rho^*)$ of a unique solution $\bar \rho$ of \eqref{eq:ann_rho} for $\ell = 0$. By the smooth dependency of the solution of a differential equation upon initial conditions and parameters, we find that $\rho_n \to \bar \rho$ uniformly on all compact subsets of $(R_0,R_1)$. Since $\rho_n$ and $\bar \rho$ are uniformly bounded thanks to \eqref{eq:estim_rho*}, the dominated convergence theorem shows that 
$$ 1 = 2 \pi \int_{R_0}^{R_1} \rho_n(r) \, r \, dr \to 2 \pi \int_{R_0}^{R_1} \bar \rho(r) \, r \, dr \quad \textrm{ as } \quad n \to \infty. $$
Hence $\bar \rho$ solves Eq. \eqref{eq:ann_rho} (with $\ell = 0$) and satisfies \eqref{eq:rho_normaliz}. So, it must be equal to $\rho_0$ defined above. Consequently, all subsequences converge to the same limit and we deduce that $\rho_\ell \to \rho_0$ as $\ell \to 0$ uniformly on all compact subsets of $(R_0,R_1)$. We now show that the convergence is uniform on $[R_0,R_1]$. Let $\varepsilon >0$. From \eqref{eq:estim_rho} and \eqref{eq:def_rho0}, there exists a compact set $K \subset (R_0,R_1)$ such that $|\rho_\ell(r)| \leq \varepsilon/2$ for all $r \in (R_0,R_1) \setminus K$ and all $\ell \geq 0$. Hence $|\rho_\ell - \rho_0| \leq \varepsilon$ on that set. By the uniform convergence of $\rho_\ell$ to $\rho_0$ as $\ell \to 0$ on $K$, we get $\limsup_{\ell \to 0} (\sup_{[R_0,R_1]} |\rho_\ell - \rho_0|) \leq \varepsilon$, which shows that $\sup_{[R_0,R_1]} |\rho_\ell - \rho_0| \to 0$. 

It follows that $v_\ell \to v_0$ as $\ell \to 0$ uniformly on $[R_0,R_1]$, with 
\begin{equation} 
v_0(r) = \frac{\rho_0(r)}{r} = \frac{1}{{\mathcal I}_1} \frac{e^{- \kappa' V(r)}}{r}, \quad \forall r \in (R_0,R_1). 
\label{eq:def_v0}
\end{equation}
Hence, the function $\ell \mapsto v_\ell^* = \max_{[R_0,R_1]} v_\ell$ belongs to $C^0([0,\infty))$.  From \eqref{eq:def_v0} and \eqref{eq:annul_cond}, we have 
$$ v_0^* \leq \frac{1}{{\mathcal I}_1 R_0} \leq \frac{c_1}{|bm|}. $$
On the other hand, thanks to \eqref{eq:defv}, we have $ v_\ell^* \geq v_\ell(R_1) = \ell R_1$. Let $ \ell_0 = \frac{2c_1}{R_1 |bm|}$. Then, $v_{\ell_0}^* > \frac{c_1}{|bm|}$. By the continuity of the function $\ell \mapsto v_\ell^*$, there exists $\ell^* \in [0,\ell_0]$ such that $v_\ell^* < \frac{c_1}{|bm|}$ for all $\ell \in [0,\ell^*)$ and that $v_{\ell^*}^* = \frac{c_1}{|bm|}$. Thus, for any $\ell \in [0,\ell^*]$, there exists a unique $v_\ell$ satisfying the constraint \eqref{eq:constraint_vell} and consequently (thanks to \eqref{eq:defv}), a unique $u_\vartheta$ satisfying \eqref{eq:constraint_utheta}. 

The end of the proof, namely the determination of $u_r$ and of $\beta$, is similar to the strip case (see Section \ref{sec_strip_proofs}). We have $u_r = \sigma (1 - u_\vartheta^2)^{-1/2}$ with  $\sigma \in \{ \pm 1 \}$, where, for $\ell < \ell^*$,  $\sigma$ is constant throughout $[R_0,R_1]$, and for $\ell = \ell^*$, $\sigma$ changes sign at each point $r^*$ where $|u_\vartheta(r^*)|=1$. In either case, there are exactly two opposite solutions for $u_r$. Once $u_r$ is determined, $\beta$ is found by integrating \eqref{eq:zero_current} and is unique provided $\beta(R^*) = 0$ is imposed. This ends the proof. \endproof

\begin{figure}[ht!]
\centering

\includegraphics[trim={4cm 15cm 5.5cm 4.5cm},clip,height= 6.5cm]{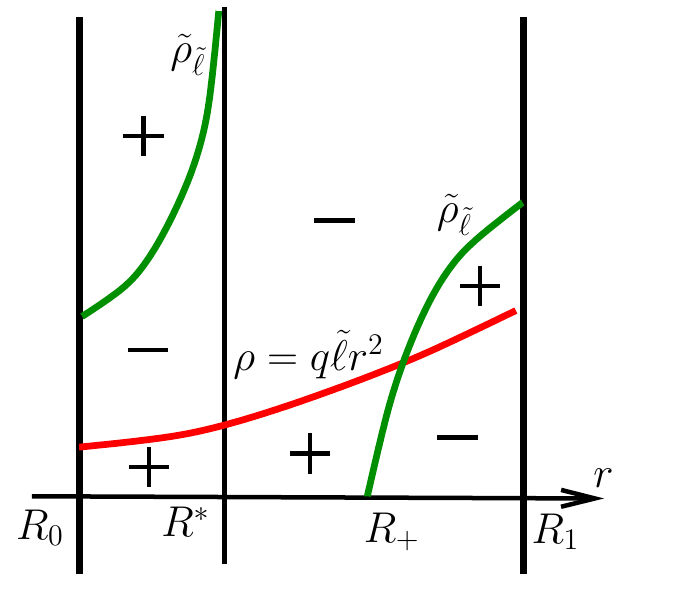}

\caption{Sign of the function $(r,\rho) \mapsto \tilde G_{\tilde \ell} (r,\rho)$ as a function of $(r,\rho) \in [R_0,R_1] \times [0,\infty)$. The graph of the function $\rho \mapsto \hat \rho_{\tilde \ell}(r)$ is depicted in green: it is the loci of points where solutions of \eqref{eq:tilGtilell} have vanishing derivatives. The graph of the function $\rho \mapsto q \tilde \ell r^2$ is represented in red: it is the loci of points where solutions of \eqref{eq:tilGtilell} have infinite derivatives.}
\label{fig:annulus_tilGtilell}
\end{figure}

Finally, we justify why we cannot reproduce this proof in the case $\ell <0$. Suppose $\ell <0$ and like in the strip geometry case, introduce $\tilde \ell = - \ell$ and 
$$ \tilde F_{\tilde \ell} (r,\rho) =  F_\ell(r,\rho) = - \kappa' V'(r) \big( \rho - \hat \rho_{\tilde \ell}(r) \big), $$
with 
\begin{equation}
\hat \rho_{\tilde \ell}(r)  = \tilde \rho_\ell(r) = - \tilde \ell r \Big( \frac{a}{\kappa' V'(r)} -r \Big). 
\label{eq:hatrho}
\end{equation}
Then, the differential equation \eqref{eq:ann_rho} is written
\begin{equation}
\rho' = \frac{\rho}{\rho - q \tilde \ell r^2} \tilde F_{\tilde \ell} (r,\rho) =: \tilde G_{\tilde \ell} (r,\rho). 
\label{eq:tilGtilell}
\end{equation}
In Fig. \ref{fig:annulus_tilGtilell}, the sign of the function $(r,\rho) \mapsto \tilde G_{\tilde \ell} (r,\rho)$ is represented. Compared to Fig. \ref{subfig:annulus_rho_construct}, we see that the phase portrait of the differential equation \eqref{eq:tilGtilell} is much more complicated than in the case $\ell >0$. We also have to account for the fact that the red curve of Fig. \ref{fig:annulus_tilGtilell} is potentially a set of singularities for $\rho$ which might prevent some solutions to be defined in the entire interval $[R_0,R_1]$. In view of this situation it seems improbable to achieve as neat a result as Prop.~\ref{prop:rotat_sym} and we'd rather use numerical simulations to find solutions in the case $\ell <0$. This will be performed in future work.

\section{Numerical simulations in strip geometry}
\label{sec:numerics}

\subsection{Setting} 
As a numerical illustration, we have simulated the travelling solution in a strip geometry corresponding to $\ell=\ell^*>0$ and $u_1 \equiv u_1^+$ (see Fig.~\ref{subfig:strip_u1u2_l=lV}). We have restricted ourself to that particular solution since a natural requirement for the stability of a solution in the strip geometry would be $(u_1 \rho')|_{\pm 1/2} >0$ (See also Section \ref{sec:conclusion}). Although further rigorous mathematical analyses would be required to confirm this conjecture, we have observed numerically that it indeed corresponds to the most stable situation in comparison to all the other classes of solutions. 

The numerical code as well as the supplementary videos (numbered 1 to 3) corresponding to the results described below are freely available at 
\begin{center}
\url{https://github.com/antoinediez/Swarmalators}. 
\end{center}

\begin{figure}[ht!]
\centering

\includegraphics[width=0.8\textwidth]{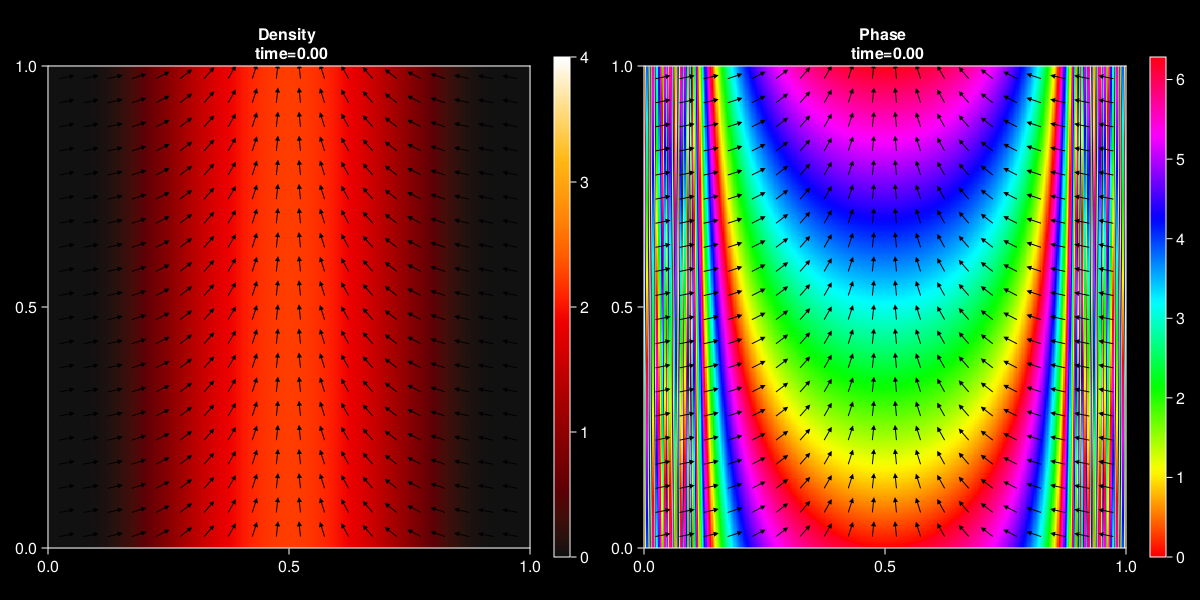}

\caption{Initial condition of the numerical simulations. The solution Eq.~\eqref{eq:express_rho} is discretized on grid of size $400\times400$. The solution is depicted within a period along the $x_2$-direction. The left panel shows the heatmap of the density $\rho$ and the velocity vector $u$. The right panel shows the value of the phase angle $\alpha$ between 0 and $2\pi$ and the velocity vector $u$.}
\label{fig:initial_conditon}
\end{figure}

This code is based on  a custom finite volume scheme with adaptive time stepping written in the \texttt{Julia} programming language and adapted from our previous work \cite{degond2022topological}. The initial condition is computed exactly, given by Eq.~\eqref{eq:express_rho}, and depicted in Fig.~\ref{fig:initial_conditon}. To do so, the implicit function $G$ and the value $\ell^*$ are computed using the nonlinear solver \texttt{NLsolve.jl} \cite{patrick_kofod_mogensen_2020_4404703}. We use the confinement potential $V(x) = ((\frac{1}{2}-x)(\frac{1}{2}+x))^{-\beta} - 4^\beta$ with $\beta=0.75$. We also provide a similar implementation to compute the other implicit functions $G_i$ corresponding to the other classes of solutions although we do not study them in the present article and postpone a more in-depth numerical study to a later work. 

\subsection{Results} 

\begin{figure}[htbp]
\centering

\subfloat[]{\includegraphics[width=0.64\textwidth]{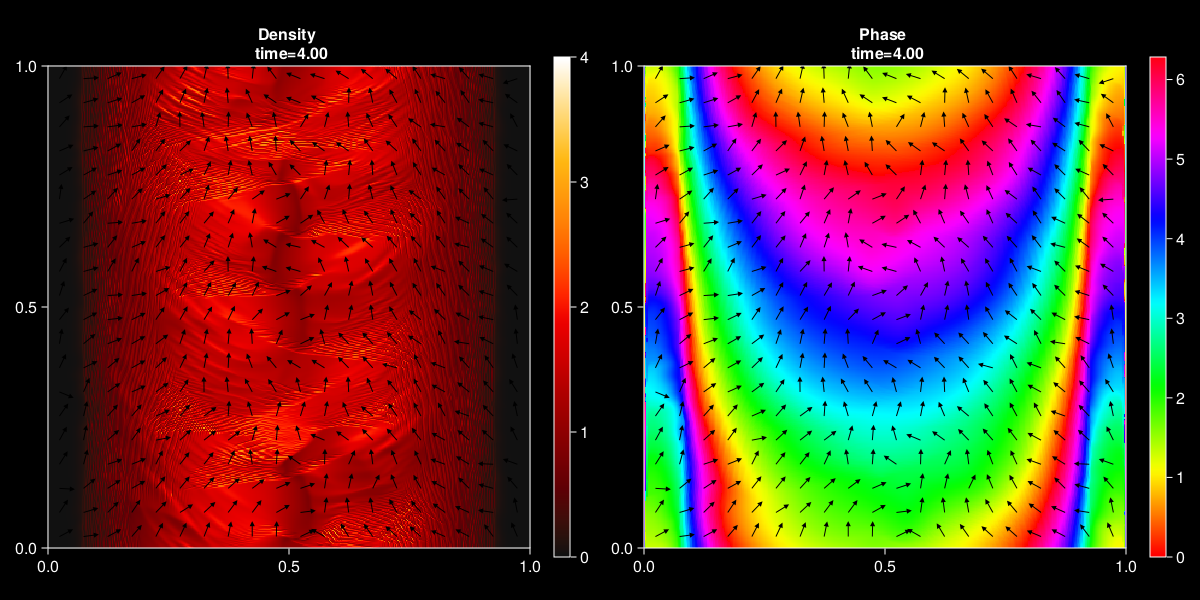}\label{fig:no_noise}}

\subfloat[]{\includegraphics[width=0.64\textwidth]{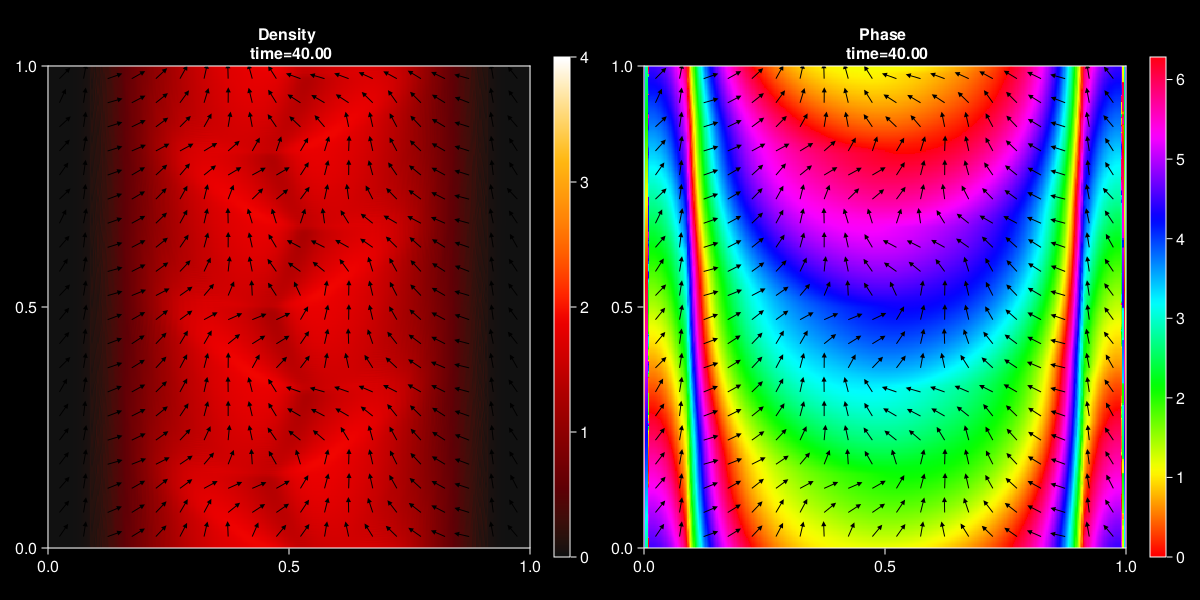}\label{fig:low_noise}}

\subfloat[]{\includegraphics[width=0.64\textwidth]{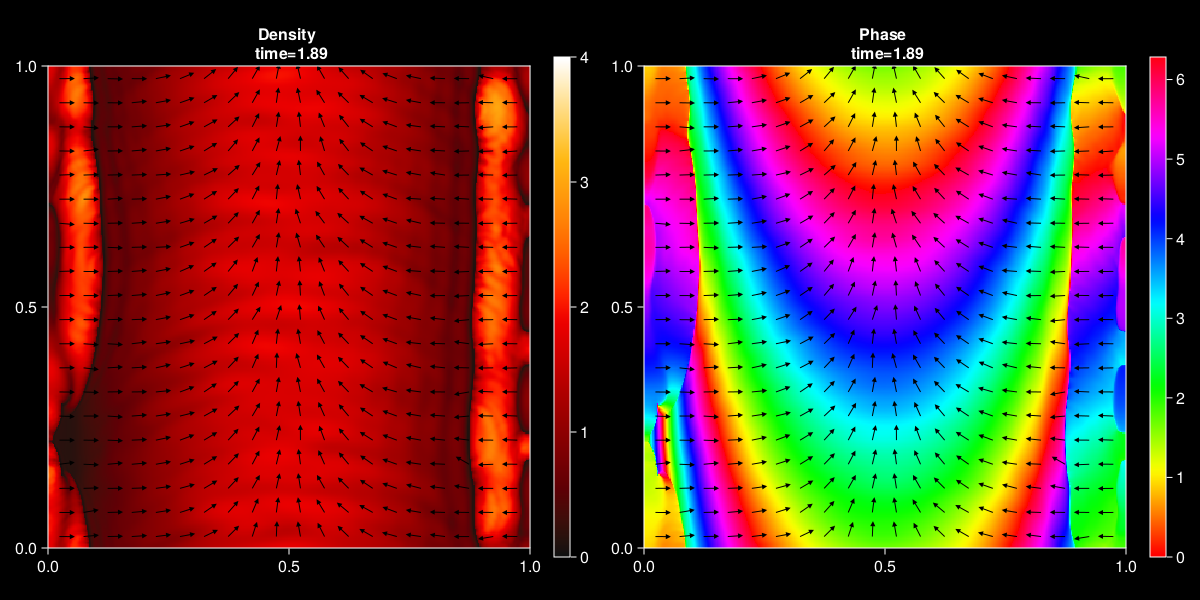}\label{fig:large_noise}}

\caption{Solutions of the NHS system (a) and of the NS system (b) \& (c) in strip geometry (see caption of Fig \ref{fig:initial_conditon} for significance of colors and arrows). The time-dependent behavior is more clearly seen in the supplementary Videos available at \url{https://github.com/antoinediez/Swarmalators}. (a) NSH system with $b'=b=-0.03267$ and $\Theta'=0$ at time $t=4.00$. Shock-like structures are quickly formed. (b) SH system at low noise ($b'=b=-0.03267$ and $\Theta'=0.01 \, b$) at time $t=40.0$. The travelling-wave solution is meta-stable and switches at time $t \sim 10.0$ towards another more stable solution where the travelling-wave behaviour is modulated by a periodic oscillation along the $x_1$ direction. (c) SH system at large noise $b=b'=-0.0308$ and $\Theta'=0.07b$ at time $t=1.89$. An instability develops from the boundary and the solution can only be simulated for very short time. }
\label{fig:solutions}
\end{figure}

Similarly to the results observed in \cite{degond2022topological} in a doubly-periodic setting, the numerical simulation of the NSH system Eqs.~\eqref{eq:fl_rho_sm}-\eqref{eq:fl_al_sm} quickly produces non-smooth solutions and shock-like structures (see Fig.~\ref{fig:no_noise} and the supplementary Video~1). A simple way to generate a smooth solution is to rather consider the SH system Eqs.~\eqref{eq:fl_rho}-\eqref{eq:fl_al} (which, at the particle level, would correspond to adding noise in the phase synchronization equation). Note however that, in contrast to the doubly-periodic setting, the proof of existence of a travelling-wave solution shown in Proposition~\ref{prop:strip} does not directly extend to the SH system so we cannot guarantee that the initial condition actually corresponds to a travelling-wave solution when $\Theta'$ and $b'$ significantly deviate from $0$ and $b$ respectively. In the results below, we have chosen $b'=b$ and $\Theta'$ equal to 1\% of $b$. In this regime, we expect the solution to be sufficiently close to the theoretical NSH travelling-wave solution, but possibly with a different travelling-wave speed. This is confirmed by the results shown in the supplementary Video~2 where a steady travelling-wave close to the theoretical solution is observed during about 10 units of time. However, this solution seems only meta-stable and after 10 units of time the solution is destabilized towards another apparently more stable solution (for the remaining 30 units of time of the simulation) where the travelling-wave behavior along the $x_2$ axis is modulated by an periodic oscillation along the $x_1$ direction. The corresponding $(\rho,u)$ is not independent of $x_2$ and $t$ any more, as shown in Fig.~\ref{fig:low_noise}.

Note that, as explained in Section~\ref{subsec:ell>0}, the phase vector associated the theoretical solution makes an infinite number of rotations as one approaches the boundary $|x_1|\to\frac{1}{2}$. This behavior cannot be preserved numerically (due to the finite size mesh) but it seems to be sufficiently mitigated by the fact that the density $\rho$ tends to zero at the boundary. It is however the main source of numerical instability as soon as more noise is added, i.~e.~when $b'$ and $\Theta'$ are too far from $b$ and $0$ respectively (see Fig.~\ref{fig:large_noise} and the supplementary Video~3 when $b=b'$ and $\Theta'$ is equal to about $7\%$ of $b$). 

\begin{remark}
As explained in \cite{degond2022topological}, the parameters $b$, $b'$ and $\Theta'$ originate from the underlying particle system and have an explicit expression in terms of the parameters of the particle model. Their detailed expressions are given by \cite[Eqs. (B.5)-(B.35)-(B.37)]{degond2022topological}. In short, these formulas depend on three components: first the magnitude $\gamma$ of the attraction-repulsion force which comes from the phase difference between two particles, secondly the concentration parameter $\kappa_v$ defining the level of noise competing with velocity alignment between neighbouring particles and finally, the concentration parameter $\kappa_\varphi$ which defines the level of noise competing with the phase synchronization between neighboring particles. With the formulas defined in \cite{degond2022topological}, the low noise case considered here corresponds to a concentration parameter $\kappa_\varphi=100$ and the large noise case to $\kappa_\varphi=15$ and in all cases $\gamma=0.033$ and $\kappa_v=1.4$. 
\end{remark}

\setcounter{equation}{0}
\section{Conclusion and discussion}
\label{sec:conclusion}

In this paper, we have studied the macroscopic swarmalator model derived in \cite{degond2022topological} and shown the existence of new smooth topological travelling-wave solutions of this model in strip and annular geometries. In the strip geometry case, we have given a complete characterization of the conditions under which such travelling-wave solutions exist thanks to the explicit integrability of the system. By contrast, in the annular geometry case, no quasi-explicit formula is available. As a consequence, we have only proved the existence of a subclass of solutions (those for which $\ell >0$) and given sufficient conditions for existence or non-existence of solutions of this class. 

One striking feature of the obtained results is that there are many possible travelling-wave solutions of the system in these geometries. However, probably, a much smaller subset of such solutions are stable. From preliminary numerical simulations, it seems that a requirement for stability is that $(u_1 \rho')|_{\pm 1/2} >0$ in the strip geometry case and $(u_r \rho')|_{R_0,R_1} >0$ in the annular geometry case. But more systematic simulations as well as rigorous mathematical results are required to confirm this conjecture. Assuming though that this stability condition is true, it only selects a very small subclass of solutions. For instance, in the strip geometry case and if $\ell >0$, it only selects the solution of Fig.~\ref{subfig:strip_u1u2_l=lV} corresponding to $u^+$, which is associated to the extremal value $\ell^*$ of the interval of admissible $\ell$. 

Of course, confirmation of these facts requires exhaustive numerical simulations which will be reported in forthcoming work. Thanks to the experience gained from \cite{degond2022topological} and in view of the preliminary numerical simulations reported here, we may attempt some predictions. First, it is unlikely that any of these travelling-wave solutions be stable for the particle system corresponding to the NSH system \eqref{eq:fl_rho_sm}-\eqref{eq:fl_al_sm}. More likely, nonsmooth travelling solutions will be generated as it was observed in the doubly-periodic geometry of \cite{degond2022topological} and here in the strip geometry case with preliminary numerical simulations. The introduction of a small amount of noise in the phase equation at the particle level, which results in the hydrodynamic system being the SH system \eqref{eq:fl_rho}-\eqref{eq:fl_al} will probably cure this deficiency, and we will most likely observe some of the travelling-wave solutions exhibited here. Whether we will observe all of them or only a subclass of those is an open question. Moreover, the numerical simulations suggest the existence of other travelling-wave solutions which are non-constant along the strip axis. Similar solutions may also exist in the annular geometry although it remains to be confirmed. Finally, there might be a discrepancy between the stability of these travelling-wave solutions between the particle system and the hydrodynamic one as was observed in the doubly-periodic geometry in~\cite{degond2022topological}. In any case, the present work opens many new questions that will require further investigations.


\end{document}